\documentclass{iopart}
\usepackage{cite}
\usepackage{comment}
\usepackage[dvipdfmx]{graphicx}
\newcommand{\mathsym}[1]{{}}

\newcommand{\si}{$\rm{i}$}
\newcommand{\sii}{$\rm{i\hspace{-.1em}i\hspace{-.1em}}$}
\newcommand{\siii}{$\rm{i\hspace{-.1em}i\hspace{-.1em}i}$}
\newcommand{\siv}{$\rm{i\hspace{-.1em}v}$}
\newcommand{\sv}{$\rm{\hspace{-.1em}v}$}
\newcommand{\svi}{$\rm{\hspace{-.1em}v}$}
\newcommand{\1}{\mbox{1}\hspace{-0.25em}\mbox{l}}

\usepackage{ulem}
\usepackage{color}

\usepackage{iopams}  

\begin{document}

\title[Exact regimes of collapsed and extra two-string solutions]{Exact regimes of collapsed and extra two-string solutions in the two down-spin sector of the spin-1/2 massive XXZ spin chain}

\author{Takashi Imoto${}^{1}$, Jun Sato,${}^{2}$ and Tetsuo Deguchi${}^{3}$}
\address{${}^{1}$Department of Basic Science, University of Tokyo, 3-8-1 Komaba, Meguro, Tokyo 153-8902, Japan\\
${}^{2}$Research Center for Advanced Science and Technology, University of Tokyo, 4-6-1 Komaba, Meguro-ku, Tokyo 153-8904, Japan
\\
${}^{3}$Department of Physics, Ochanomizu University, 2-1-1 Ohtsuka, Bunkyo-ku, Tokyo 112-8610, Japan}

\ead{deguchi@phys.ocha.ac.jp; t-imoto@g.ecc.u-tokyo.ac.jp}
\vspace{10pt}
\begin{indented}
\item[]\today
\end{indented}

\begin{abstract}
We derive exactly the number of  complex solutions with two down-spins 
in the massive regime of the periodic spin-1/2 XXZ spin chain of $N$ sites.  Here we remark that every solution of the Bethe ansatz equations is characterized by a set of quantum numbers. We derive them analytically for all the complex solutions  in the sector, which we call two-string solutions. We show that in a region of $N$ and $\Delta$ the number of two-string solutions  
is by two larger  than the number due to the string hypothesis,  i.e., an extra pair of two-strings  appears. We  determine it exactly and also such regions where $m$ two-string solutions collapse for any positive integers $m$. We illustrate the extra and standard two-string solutions numerically. In the sector we show that the string deviations are exponentially small with respect to $N$ if $N$ is large.  We argue that for any finite solution of the spin-1/2 XXX chain there is such a solution of the spin-1/2 XXZ chain that has the same quantum numbers in common with the XXX solution.  
\end{abstract}

\vspace{2pc}

\section{Introduction}
In the study of quantum many-body systems, several numerical techniques 
 have been developed to evaluate their eigenvalues, eigenvectors,  and physical quantities, at least 
approximately. In order to examine the accuracy of such numerical techniques, 
exact solutions should be useful. However, the systems for which we obtain all eigenvectors and eigenvalues exactly are only a few. Some quantum integrable systems in one dimension are part of them. We call  models which satisfy the Yang-Baxter relation  quantum integrable systems.
The Bethe ansatz is useful to derive systematically eigenvectors and eigenvalues for them\cite{Bethe}. Here the Bethe ansatz equations play a central role \cite{Baxter,FT}. 
From a solution of the Bethe ansatz equations  we obtain an eigenvalue and its eigenvector.

The spin-1/2 anisotropic quantum Heisenberg spin chain, i.e.,   
the spin-1/2 XXZ spin chain under the periodic boundary conditions is one of the 
fundamental quantum integrable systems in one dimension.
The Hamiltonian of this model is given by
\begin{eqnarray}
H_{XXZ}=\frac{1}{4}\sum_{j=1}^{N}\biggl(\sigma_{j}^{x}\sigma_{j+1}^{x}+\sigma_{j}^{y}\sigma_{j+1}^{y}+\Delta(\sigma_{j}^{z}\sigma_{j+1}^{z}-\1)\biggr)
\end{eqnarray}
where $\sigma_{j}^{a}(a=x,y,z)$ are the Pauli matrices defined on the $j$th site, $\Delta$ denotes the anisotropic parameter, and $N$ is the site number. 
If $| \Delta | > 1$, the spectrum has a gap, i.e., the energy difference between the first excited state and the ground state remains nonzero and finite as the system size becomes infinitely large, while if $|\Delta| \le 1$ it is gapless or massless,  i.e., the energy gap vanishes in the thermodynamic limit.    
In the $M$ down-spin sector with rapidities $\lambda_{1}, \lambda_{2}, \cdots$, 
and $\lambda_{M}$, 
the Bethe-ansatz equations of the spin-1/2 XXZ Heisenberg spin chain are given by 
\begin{eqnarray}
\Biggl(\frac{\phi(\lambda_{j}+i\zeta/2)}{\phi(\lambda_{j}-i\zeta/2)}\Biggr)^{N}=\prod_{k\neq j,k=1}^{M}\frac{\phi(\lambda_{j}-\lambda_{k}+i\zeta)}{\phi(\lambda_{j}-\lambda_{k}-i\zeta)}\ \ \ (j=1,2,\cdots ,M).\label{eq:BAE0}
\end{eqnarray}
For $\Delta=1$ (the XXX case) we take $\phi(\lambda)=\lambda$ and $\zeta=1$. For  $-1<\Delta<1$ (the massless regime of  the XXZ spin chain) we take $\phi(\lambda)=\sinh\lambda$ and $\Delta=\cos\zeta$. For $\Delta>1$ case (the massive regime of the XXZ spin chain)  we take $\phi(\lambda)=\sin\lambda$ and $\Delta=\cosh\zeta$.

We introduce quantum numbers  
in order to solve the Bethe ansatz equations (\ref{eq:BAE0}) numerically. 
We express the logarithms of both hand sides of eqs. (\ref{eq:BAE0}) 
in terms of an analytic continuation of the arctangent function  denoted by $\widehat{\tan^{-1}}(x)$,  which will be defined shortly. 
For $\Delta>1$ we have
\begin{eqnarray}
2\widehat{\tan^{-1}}\biggl(\frac{\tan{\lambda_{i}}}{\tanh(\zeta/2)}\biggr)=\frac{2\pi}{N}J_{i}+\frac{1}{N}\sum_{k=1}^{M}2\widehat{\tan^{-1}}\biggl(\frac{\tan(\lambda_{i}-\lambda_{k})}{\tanh{\zeta}}\biggr)  \, ,  \label{eq:logBAE0}  \\
J_{i}\equiv\frac{1}{2}(N-M+1)\ \ (\mbox{mod} \, 1)  \quad  \mbox{for} \, \,  i = 1, 2,\cdots, M. 
\label{eq:QN}
\end{eqnarray}
We call $J_{i}$ the Bethe quantum numbers.  The conditions (\ref{eq:QN}) with 
mod 1  mean that they are half-integers if $N-M$ is even, 
and integers  if $N-M$ is odd.

It seems as if it is easy to solve the Bethe ansatz equations (\ref{eq:logBAE0}) 
numerically and evaluate physical quantities for large systems. 
However,  it is not trivial to derive solutions to eqs. (\ref{eq:logBAE0}) numerically 
not only in the thermodynamic limit but also in finite-size systems particularly 
for excited states with complex solutions. 
In fact, it is not known how to specify the Bethe quantum numbers for 
an arbitrarily given eigenstate.  
For instance, the ground-state energy of the antiferromagnetic spin-1/2 XXX spin chain   
was calculated by Hulth\'{e}n successfully \cite{Hul}, partially because 
every ground-state rapidity is real \cite{YY}.
However, solutions of eqs. (\ref{eq:logBAE0})  for excited states 
may be complex. There is a set of numerical assumptions on the forms of complex solutions, 
which we call the string hypothesis \cite{TK1,TK}.  
Here we remark that there are also combinatorial approaches for classifying solutions of the 
Bethe ansatz equations \cite{KKR,KR,KS1,DG2,GD2}. 

Let us consider the two down-spin sector in the spin-1/2 XXX spin chain or  that of  
the massive spin-1/2 XXZ spin chain.  
There are both real and complex solutions for the Bethe-ansatz equations.  
 A pair of complex rapidities, which we call a two-string,  
is expressed by two real parameters such as string center $x$ and string deviation $\delta$ \cite{Vl1}. 
For the massive XXZ case where $\Delta = \cosh \zeta$ and  $\delta > - \zeta/2$, we have 
\begin{eqnarray}
\lambda_{1}=x+\frac{i}{2}\zeta+i\delta \, ,  \nonumber \\
\lambda_{2}=x-\frac{i}{2}\zeta-i\delta .  \label{2string}
\end{eqnarray}
Some of two-string solutions predicted by the string-hypothesis 
become real solutions if  the site number $N$ is large, for the spin-1/2 XXX spin chain 
in the two down-spin sector  \cite{EKS}.  We call it the collapse of two-string solutions to real ones. The critical number $N_c$ such that a collapsed two-string solution exists 
for $N>N_c$ is evaluated as $21.86$ in the spin-1/2 XXX spin chain \cite{EKS,HC}.  
The collapse is also numerically investigated \cite{FKT}.  
The number of collapsed solutions is rigorously obtained 
for any given number $N$ of the spin-1/2 XXX spin chain by deriving  the Bethe quantum numbers  
\cite{DG1}. 

The Bethe ansatz equations for the  spin-1/2 XXX and XXZ chains may 
have solutions containing a pair of pure imaginary rapidities $(i\zeta/2,-i\zeta/2$). 
We call them singular solutions \cite{AV,Vl2,NW1,NW2,HNS1,HNS2,GD1,Si,BMSZ,KS2,D1,Vl1}.  
They make some factors of eqs. (\ref{eq:BAE0}) indefinite. It is not straightforward to        
show that they correspond to eigenvectors, However, 
for instance, we can show it by deriving the corresponding Bethe quantum numbers \cite{DG1}. 

It is therefore fundamental to derive the Bethe quantum numbers for solutions of  
the Bethe ansatz equations when we evaluate them numerically in some sector.  At the XXX point, 
all Bethe quantum numbers both for real and complex solutions are derived rigorously  in the two down-spin sector \cite{DG1}.  In the spin-1/2 XXZ chain in the two down-spin sector, the completeness has been proved rigorously \cite{KE}. However, the Bethe quantum numbers have not been discussed, yet. Here we mention earlier studies on complex solutions of the spin-1/2 XXZ spin chain\cite{Wo,BVV,FM1,FM2}.

\begin{figure}[htbp]
\begin{center}
 \includegraphics[width=8cm,clip]{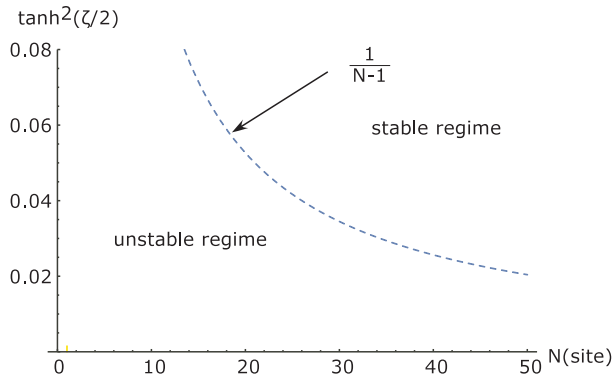}
\end{center}
 \caption{Stable and unstable regimes. Vertical axis denotes the value of 
$\tanh^2(\zeta/2)$, while the horizontal axis the number of sites $N$. 
The dashed line shows the boundary of the two regimes given by 
the graph of $ \tanh^2(\zeta/2) = 1/(N-1)$. 
}
\label{fig:stable-unstable} 
\end{figure}

\begin{figure}[htbp]
\begin{center}
 \includegraphics[clip,width=10cm]{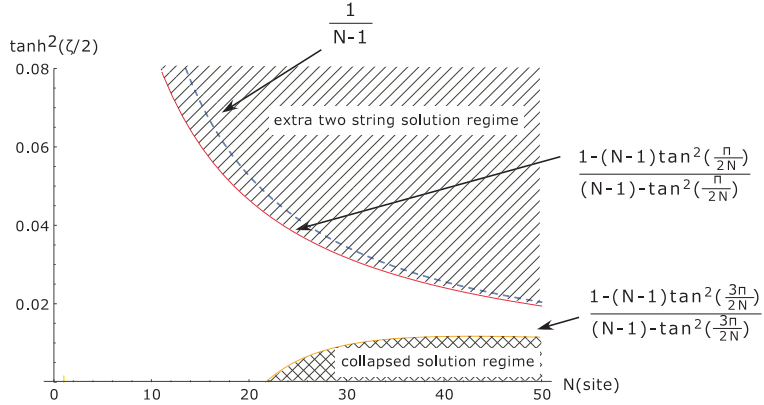}
\end{center}
 \caption{Extra pair of 2-string solutions appears in the area indicated by oblique lines, which is above the real red line. The dashed line shows the boundary between the stable regime and the unstable regime. The region of $k$ missing 2-string solutions 
is crosshatched.  }
 \label{fig:extra-sol} 
\end{figure}

\begin{figure}[htbp]
\begin{center}
 \includegraphics[clip,width=10cm]{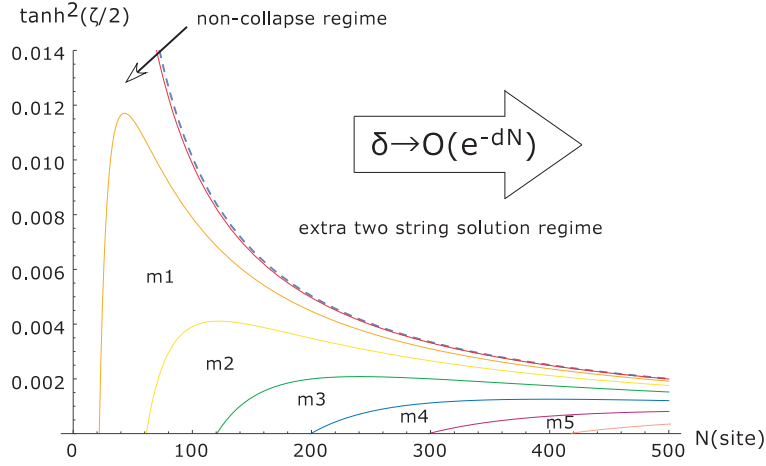}
\end{center}
 \caption{Stable regime is above the dashed line, while unstable regime is below it.    
In the stable regime the string deviations decrease exponentially with respect to $N$. 
Symbols m$k$ such as m$1$, m$2$, $\cdots$, denote the regions of $k$ missing 
2-string solutions for $k=1, 2, \ldots$.
Vertical axis shows the value of  $\tanh^{2}(\zeta/2)$ and 
horizontal axis  the number of sites $N$. 
 The XXZ anisotropy $\Delta$ is given by $\Delta=\cosh\zeta$. }
    \label{fig:missing-sol} 
\end{figure}

In this paper,  we derive the number of complex solutions in the two down-spin sector analytically  by showing all the Bethe quantum numbers  
for the massive regime of the spin-1/2 XXZ spin chain. 
There are several results:  
(\si) The number of two-string solutions depends on $N$ and $\Delta$. We show  that it is by two larger than the number given  by the string hypothesis in a wide region of $\Delta$ and $N$.
We call them an extra pair of two-string solutions. 
(\sii) Let us introduce $N_{\zeta}$ for  positive values of $\zeta$ by 
\begin{equation} 
N_{\zeta} = 1 + 1/\tanh^2(\zeta/2) . \label{eq:Nzeta}
\end{equation} 
We call   such a regime of $\Delta$ and $N$ satisfying $N > N_{\zeta}$ for $\zeta > 0$ the {\it stable regime}.  It  is above the dashed lines in figure \ref{fig:stable-unstable}.  We show that  in the stable regime 
all the two-string solutions predicted by the string hypothesis exist 
and the string deviations $\delta$ are exponentially small with respect to  $N$.  
Furthermore, we show that an extra pair of two-string solutions exists throughout the stable regime. We conclude that  two-string solutions are stable in the stable regime. In short,  two-string solutions are stable for two down-spins 
if the site number $N$ or the XXZ anisotropy $\Delta$ is large enough.   
Here we remark that the criterion: $N > N_{\zeta}$ is equivalent to 
the condition: $(N-1) \tanh^2(\zeta/2) > 1$.  It is also expressed 
as $\tanh^2(\zeta/2) > 1/(N-1)$. 
(\siii) We call such a regime of $\Delta$ and $N$ 
satisfying $N < N_{\zeta}$ for $\zeta > 0$ the {\it unstable regime}. 
It is below the dashed lines in each of figures \ref{fig:stable-unstable}, \ref{fig:extra-sol} and \ref{fig:missing-sol}. 
In the unstable regime, for any given positive integer $m$, 
 we determine  such a region of $N$ and $\Delta$ where the collapse of 
$m$ two-string solutions occurs.  Thus, two-string solutions are 
possibly unstable in the unstable regime.   
(\siv) We rigorously show that the string deviations $\delta$ are exponentially small  with respect to $N$ in the stable regime by an analytic method.  
We illustrate it explicitly with numerical solutions of the Bethe ansatz equations.  Furthermore,  we present numerical solutions of an extra pair of two strings  
for $N=12$. We thus confirm the existence of extra two-string solutions numerically. 
(\sv) We argue that if there is a finite-valued solution 
of the Bethe-ansatz equations in the XXX spin chain, 
then there exists a solution of the XXZ spin chain that has the same 
Bethe quantum numbers as the XXX solution and also that 
the XXZ solution converges to the XXX solution 
in the isotropic limit where we send $\zeta$ to zero. 
We call it the XXX/XXZ correspondence of the Bethe quantum numbers.

Let us derive the logarithmic form of the Bethe ansatz equations from eqs. (\ref{eq:BAE0}) more explicitly. For simplicity, we assume that all rapidities $\lambda_j$ are real.  
We remark that each rapidity $\lambda_j$ has the period $\pi$
in the Bethe ansatz equations (\ref{eq:BAE0}), and hence we search solutions satisfying $- \pi/2 < \lambda_j \le \pi/2$ for all $j$. 
Applying the relation $(x+i/n)/(x-i/n)=(-1) (1-i n x)/(1 + i n x)$ and 
eqs. (\ref{eq:A1}) in Appendix A to the right-hand side of eqs. (\ref{eq:BAE0}), we have 
\begin{eqnarray}
\frac{\phi(\lambda_{j} - \lambda_k + i \zeta)}
                 {\phi(\lambda_{j}- \lambda_k - i \zeta)} &
= \frac{\tan(\lambda_{j}-\lambda_k)/\tanh(\zeta) + i }
                   {\tan(\lambda_{j}-\lambda_k)/\tanh(\zeta) - i} 
\nonumber\\
&=(-1)  \exp \biggl[- i 2\tan^{-1}\biggl(\frac{\tan(\lambda_{j}-\lambda_k)}{\tanh(\zeta)} \biggr)\biggr]. \label{eq:logrhs}
\end{eqnarray}
Here we take the principal branch for 
$\tan^{-1}(x)$ such as $-\pi/2 < \tan^{-1}(x) < \pi/2$ for $-\infty < x < \infty$. 
In the right hand side of (\ref{eq:logrhs}) the difference $\lambda_j - \lambda_k$ may be 
greater than $\pi/2$. In order to make the phase factor continuous at 
$\lambda_j - \lambda_k= \pi/2$ Takahashi introduced the following function 
for real variable $\lambda$ \cite{TK}, 
\begin{equation}
\theta_n(\lambda; \zeta) = 
2\tan^{-1}\biggl(\frac{\tan \lambda}{\tanh(n \zeta/2)}\biggr)
+ 2 \pi \left[ {\frac {2 \lambda + \pi} {2 \pi}} \right] \, . 
\end{equation}
Here we denote by the symbol $[x]$ the greatest integer that is not larger than $x$. 
In order to search real solutions numerically by the recursive method, 
it is convenient to make the phase factor continuous at  $\lambda_j- \lambda_k=\pi/2$. 
In the paper, in order to derive complex solutions 
we analytically continue the functions $\theta_n(\lambda; \zeta) $ 
through the two branches of the logarithmic function 
\begin{equation}
2 \widehat{ \tan^{-1} }\left( x + iy  \right)
= {\frac 1 i} \left( \log^{(+)}(1 +i (x + iy)) - \log^{(s)}(1 - i (x + iy)) \right) \, .  
\label{eq:arctangent}
\end{equation}
Here the branches of the logarithmic function $\log^{(s)}(z)$ and  $\log^{(+)}(z)$ are given in eqs. (\ref{eq:A1}) with (\ref{eq:A2}) and eqs. (\ref{eq:A3}) with (\ref{eq:A4}), 
respectively.  

Let us  express  deviation $\delta$ in terms of a variable $w$ defined by   
\begin{equation}
w\equiv \tanh(\delta+\zeta/2)/\tanh(\zeta/2) . \label{eq:def-w}
\end{equation}
We consider two intervals of $w$: the interval between 0 and 1  
for narrow strings with $\delta < 0$ denoted by $(0, 1)$, 
and the interval  between 1 and $1/\tanh(\zeta/2)$
for wide strings with $\delta > 0$ denoted by $(1, 1/\tanh(\zeta/2) )$.  

By setting the imaginary part of the Bethe ansatz equations to be zero, 
we regard the center $x$ as a function of variable $w$.  We shall express  $\tan x$ as a function of $w$.  We introduce the counting function $Z_1(w)$ as a function of $w$ for the 2-string solutions in the two down-spin sector with eqs. (\ref{2string})  as follows. 
\begin{eqnarray}
2 \pi Z_1(w)  & = & 2\widehat{\tan^{-1}}\biggl(\frac{\tan{\lambda_1 }} {\tanh(\zeta/2)}\biggr)
- {\frac 1 N} 2 \widehat{\tan^{-1}} \biggl( \frac {\tan(\lambda_{1}-\lambda_{2})}{\tanh{\zeta}} \biggr) \label{eq:CF}
\end{eqnarray}
Here $\lambda_1$ and $\lambda_2$ are functions of $w$ through 
the relation of string center $x$ to $w$. 
We obtain a two-string solution $\lambda_1(w(J_1))$ and $\lambda_2(w(J_1))$  
for a given set of Bethe quantum numbers $J_1$ and $J_2$ by solving 
the following equation for $w$ 
\begin{equation} 
Z_1(w(J_1))=\frac {J_{1}} N \, .   
\end{equation}
There is another counting function $Z_2(w)$ satisfying $Z_2(w(J_1))= J_2/N$.  
However,  since their difference  is constant,  
we consider only $Z_1(w)$ and denote it by $Z(w)$.  

If the counting function $Z(w)$ increases (or decrease) monotonically 
through an interval $[w_a, w_b]$, we obtain the set of Bethe quantum numbers 
by the set of such integers or half-integers $J_k$ satisfying 
$N Z(w_a) \le  J_k  \le N Z(w_b)$ (or $N Z(w_b) \le  J_k  \le N Z(w_a)$ ).     
In the stable regime we show rigorously that 
the counting function is monotone in $w$ for large $N$.   
In the unstable regime we confirm the monotonicity numerically,   
and give analytical arguments for it. 
We thus derive exact expressions of $N$ and $\zeta$ (or $\Delta$) 
for such regions where $m$ two-string solutions are missing.

We regard the range of $w$ on which we define the counting function $Z(w)$ 
as the domain of definition for the function.

Let us  explain the string hypothesis more explicitly. 
The string hypothesis consists of two parts. ($\rm{i}$) 
The complex solutions of the Bethe ansatz equations (\ref{eq:BAE0})
assumed by the string hypothesis  for 
the spin-1/2 XXZ spin chain in the massive regime 
have in general the following form:     
 \begin{eqnarray}
 \lambda_{\alpha,j}^{n}=\lambda_{\alpha}^{n}+(n+1-2j) {\frac  {i \zeta}   2} +O(e^{-d N}),\ \ \ j=1,\cdots,n\,   \label{eq:string}
 \end{eqnarray}
where $d$ is a positive constant and 
the string center $\lambda_{\alpha}^{n}$ is given by a real number satisfied by $-\pi/2<\lambda_{\alpha}^{n}\leq\pi/2$. 
We call  a complex solution of the form (\ref{eq:string}) an $n$-string. 
If the deviations $O(e^{-d N})$ in (\ref{eq:string}) vanish, 
we call it a complete $n$-string or simply a complete string.  
The set of all solutions in the $M$ down-spin chain consists of 
$k$-strings for $k=1, 2, \ldots, $ and $M$.   
The total number of $k$-strings is denoted by $M_{k}$ for each $k$.
It is clear that we have $M=\sum_{k=1}^{\infty}kM_{k}$.
($\rm{i\hspace{-.1em}i}$)The numbers of sets of $k$-strings for $k=1, 2, \ldots$, 
 are determined by the assumptions of the string hypothesis.
If we assume this hypothesis, we can evaluate the free energy at finite temperatures at least approximately. Moreover, from the string hypothesis, thermal properties, e.g., specific heats and magnetic susceptibilities are studied \cite{TY,YT,Sch2,Sch1,Z1,ZNP}.  However,  counter examples of this hypothesis are found in the spin 1/2-XXX and the spin-1/2 massless XXZ spin chains in the two down-spin sector\cite{EKS,IP,IKPP}.

There are several physical motivations in the study of the Bethe quantum numbers 
and  the validity of  the string hypothesis for numerical solutions 
of the Bethe ansatz equations  in the two down-spin sector for the 
massive regime of the spin-1/2 XXZ spin chain.   
(\si) With exact Bethe quantum numbers being given 
we can systematically evaluate numerical solutions  of the 
logarithmic forms of the Bethe ansatz equations. 
It is fundamental to know the numbers of real and two-string solutions 
when we derive numerical solutions in some sector. 
(\sii)  With numerically exact solutions, we can derive the quantum dynamics 
in a 
finite 
system for  any given initial state in the sector. 
(\siii) In the XXX chain and the massless XXZ chain, the counter example for the string hypothesis in the two down-spin sector exist. However, 
in the massive XXZ chain it seems that any counter example to  the string hypothesis or any argument supporting it is not known.  
(\siv) Recently, the spectrum in the massive regime of the spin-1/2 
XXZ spin chain has attracted much interest.  
For instance, the low-temperature spectrum of 
correlation lengths are studied in the antiferromagnetic massive regime of the spin-1/2 XXZ spin chain \cite{DGKS1,DGKS2}. 
Since some of the results of the present paper are favorable to 
the string hypothesis,  it should be interesting to examine how far    
thermal properties obtained by assuming the string hypothesis 
should be valid.

The contents of this paper consist of the following.
In section \ref{sec:TSS}, we rigorously formulate the Bethe ansatz equations for two-string solutions in the two down-spin sector. We divide each of 
the Bethe ansatz equations into the real and imaginary parts. 
We regard the imaginary parts as  constraints on string center $x$ and 
variable $w$. 
We also derive the difference of the two Bethe quantum numbers $J_1$ and $J_2$. 
Let us call two-strings with positive string deviations: $\delta> 0$ 
{\it wide pairs} and  two-strings with negative string deviations: $\delta< 0$ 
{\it narrow pairs}. 
In section \ref{sec:monotone} we argue the monotonicity of 
the counting function $Z(w)$.  
In the case of $x>0$, for instance, 
it is monotone increasing as a function of $w$ for wide pairs 
and monotone decreasing for narrow pairs.   
In the stable regime for large $N$ we prove the monotonicity. 
In the unstable regime for narrow pairs we examine the monotonicity 
numerically, and we make it a conjecture. 
In section \ref{sec:CCS}, we count the complex solutions with the Bethe quantum numbers. We derive analytically the criterion about the collapse of two-string solutions and the existence of extra two-string solutions. 
In the stable regime for large $N$ we prove the existence of an extra two-string solutions. 
Furthermore,
we derive a singular solution and calculate its Bethe quantum numbers. 
In section \ref{sec:RSH}, we show rigorously that string deviations of 
two-string solutions become exponentially small as the number of sites $N$ increases.  the relation between the string hypothesis and the Bethe solutions. 
We show that except for narrow pairs in the unstable regime 
two-string solutions  in the two down-spin sector 
approach 
complete strings when  $\Delta>1$ and $N$ is large enough.  
In section \ref{sec:numerical} 
we show numerically that  the string deviations become exponentially small as 
the site number $N$ increases.  
We explicitly presented  the numerical estimates of extra string solutions for 
$N=12$. We have illustrated the Bethe quantum numbers with the numerical solutions for $N=12$. 
We present the numerical estimates of extra string solutions for 
$N=12$. We illustrate the Bethe quantum numbers with the numerical solutions for $N=12$. 
In section \ref{sec:XXXXXZ}, we argue in the two down-spin sector 
the XXX/XXZ correspondence of the Bethe quantum numbers: for any finite solution of the XXX spin chain  there exists a solution of the XXZ spin chain that has the same Bethe quantum 
numbers as the XXX solution.

\section{Solving the Bethe ansatz equations for two-strings with two down-spins}\label{sec:TSS}
\subsection{Real and  imaginary parts of the Bethe ansatz equations} \label{sec:divide_BAE}

Let us consider the Bethe ansatz equations in the two down-spin sector, i.e., eqs. 
(\ref{eq:logBAE0}) for $M=2$. They are given by 
\begin{eqnarray}
 2\tan^{-1}\biggl(\frac{\tan\lambda_{1}}{\tanh{\zeta/2}}\biggr) 
& =\frac{2\pi}{N}J_{1}+\frac{2}{N}\tan^{-1}\biggl(\frac{\tan(\lambda_{1}-\lambda_{2})}{\tanh\zeta}\biggr) \, ,  \label{eq:BAE11}\\
2\tan^{-1}\biggl(\frac{\tan\lambda_{2}}{\tanh{\zeta/2}}\biggr)&=\frac{2\pi}{N}J_{2}+\frac{2}{N}\tan^{-1}\biggl(\frac{\tan(\lambda_{2}-\lambda_{1})}{\tanh\zeta}\biggr)\label{eq:BAE22}.
\end{eqnarray}
We substitute the two-string form (\ref{2string}) in the Bethe ansatz equations (\ref{eq:BAE11}) and (\ref{eq:BAE22}) in the two down-spin sector as follows. 
\begin{eqnarray}
&  2\tan^{-1}\biggl(\frac{\tan(x+\frac{i}{2}\zeta+i\delta)}{\tanh{\zeta/2}} \biggr)
\nonumber \\ 
&  \quad =\frac{2\pi}{N}J_{1}+\frac{2}{N}\tan^{-1}\biggl(\frac{\tan(i(\zeta+2\delta))}
{\tanh\zeta}\biggr) , \label{eq:BAE11cp} \\
&  2\tan^{-1}\biggl(\frac{\tan(x-\frac{i}{2}\zeta-i\delta)}{\tanh{\zeta/2}} \biggr) 
\nonumber \\
&  \quad =\frac{2\pi}{N}J_{2}+\frac{2}{N}\tan^{-1}\biggl(\frac{\tan(-i(\zeta+2\delta))}{\tanh\zeta}\biggr). \label{eq:BAE22cp}
\end{eqnarray}

It is convenient to introduce variable $t$ by 
\begin{equation} 
t = \tanh \left( \zeta/2 \right). \label{eq:def-t} 
\end{equation} 
Here we recall that $w$ has been defined by $w= {\tanh{(\zeta/2+\delta)}}/{\tanh{\zeta/2}}$ in 
eq.  (\ref{eq:def-w}).  We also recall that $w> 0$ since we have the constraint 
$\delta > -\zeta/2$ due to eq. (\ref{2string}).  
If deviation $\delta$ is negative,  we have $w < 1$, while if it is positive we have 
$w >1$.  We remark that $w < 1/t$ by definition. 

We first consider the left-hand side of the Bethe ansatz equations (\ref{eq:BAE11cp}). 
Thanks to the tangent angle addition formula we have 
\begin{eqnarray}
\frac{\tan{\lambda_{1}}}{\tanh{\zeta/2}}=\frac{\tan{x}(1-\tanh^2(\frac{\zeta}{2}+\delta))+i(1+\tan^{2}{x})\tanh(\frac{\zeta}{2}+\delta)}{\tanh{\zeta/2}\Bigl(1+\tan^2{x}\tanh^{2}{(\frac{\zeta}{2}+\delta)}\Bigr)}\label{eq:BAELHS1}
\end{eqnarray}
We express the real part of eq. (\ref{eq:BAELHS1}) by $a$, and the imaginary part of (\ref{eq:BAELHS1}) by $b$ 
\begin{eqnarray}
a 
&=&\frac{\tan{x}(1-w^2 t^2 )}{ t \Bigl(1+(\tan^2{x} )w^2 t^2 \Bigr)} 
\label{eq:a} \\
b
&=&\frac{(1+\tan^{2}{x})w}{\Bigl(1+(\tan^2{x})w^2 t^2 \Bigr)} \label{eq:b}.
\end{eqnarray}
We express the left-hand side of the Bethe ansatz equations in terms of the arctangent function 
with a complex argument of $a + i b$  in 
(\ref{eq:arctangent}). 
\begin{eqnarray}
& & 2 \widehat{\tan^{-1}}\biggl(\frac{\tan(x+\frac{i}{2}\zeta+i\delta)}{\tanh{\zeta/2}}\biggr)
 =  \tan^{-1}\Bigl(\frac{a}{1-b}\Bigr) + \tan^{-1}\Bigl(\frac{a}{1+b}\Bigr) \nonumber \\ 
& & +\pi\ H(b-1)  + 2 \pi H(1-b) H(-a)   - \pi H(-b-1) \mbox{sgn}(a_{-}) \nonumber \\ 
& & +\frac{1}{2i}\log\biggl(\frac{a^2+(b-1)^2}{a^2+(b+1)^2}\biggr)\label{eq:LHSBAE}
\end{eqnarray}
where, the step function $H(y)$ is defined by
\begin{eqnarray}
H(y)=\left\{ \begin{array}{ll}
    1 & \mbox{for } y>0 \, ,  \\
    0 & \mbox{otherwise} 
\, , 
\end{array} \right.
\end{eqnarray}
and $\mbox{sgn}(y_{+})$ is expressed as  $\mbox{sgn}(y_{+})=1-2H(-y)$.
Similarly, we consider the right-hand side of the Bethe ansatz equation 
(\ref{eq:BAE11cp}) as follows.
\begin{eqnarray}
\frac{2}{N} \tan^{-1} \biggl(\frac{\tan(\lambda_{1}-\lambda_{2})}{\tanh{\zeta}}\biggr)
=\frac{2}{N}\tan^{-1} \biggl(\frac{i\tanh(\zeta+2\delta)}{\tanh{\zeta}}\biggr)
\nonumber\\
=\frac{1}{N}\biggl\{\pi H(\delta)+\frac{1}{2i}\log\Biggl(\frac{(\tanh(\zeta+2\delta)/\tanh{\zeta}-1)^2}{(\tanh(\zeta+2\delta)/\tanh{\zeta}+1)^2}\biggr)\Biggr\} 
\, . 
\label{eq:RHSBAE}
\end{eqnarray}
Here we remark that it is easy to show that $b> 0$. We also recall that 
$\zeta+ 2 \delta> 0$ since  $\delta > -\zeta/2$. 
It follows from (\ref{eq:LHSBAE}) and (\ref{eq:RHSBAE}) 
that the Bethe ansatz equation (\ref{eq:BAE11cp}) is expressed as
\begin{eqnarray}
\frac{2\pi}{N}J_{1} 
&=&\tan^{-1}\biggl(\frac{a}{1-b}\biggr)+\tan^{-1}\biggl(\frac{a}{1+b}\biggr) \nonumber \\ 
& & + \pi\biggl( H(b-1) + 2 H(1-b) H(-a)  - \frac{H(\delta)}{N}\biggr)\nonumber\\
&+&\frac{1}{2i}\log\Biggl\{\biggl(\frac{a^2+(b-1)^2}{a^2+(b+1)^2}\biggr)\biggl(\frac{(\tanh(\zeta+2\delta)/\tanh \zeta + 1)^2}{(\tanh(\zeta+2\delta)/\tanh \zeta-1)^2}\biggr)^{1/N}\Biggr\}
\nonumber \\ 
\label{eq:BAE_2}.
\end{eqnarray}
We consider the real and imaginary parts of eq.  (\ref{eq:BAE_2}), separately. 
We recall that we regard $x$  as a function of $w$ (i.e., $\delta$)  
by setting  the imaginary part of eq. (\ref{eq:BAE_2}) to be zero.
From the real part of eq. (\ref{eq:BAE11cp}) we define the counting function $Z_1(w)$ as follows.
\begin{eqnarray}
Z_1(\delta(w), x(w), \zeta) & 
\equiv \frac 1 {2 \pi} \tan^{-1}\biggl(\frac{a}{1-b}\biggr)
+\frac 1 {2 \pi} \tan^{-1}\biggl(\frac{a}{1+b}\biggr)
\nonumber \\   
& +{\frac 1 2} \biggl( H(b-1) + 2 H(1-b)H(-a) - \frac{H(\delta)}{N}\biggr) . 
\label{eq:counting1}
\end{eqnarray}
For a given Bethe quantum number $J_1$, we determine the solution $\lambda_1$ and $\lambda_2$ which corresponds to $J_1$ by evaluating $w$ such that it satisfies the following equation. 
\begin{eqnarray}
Z_1(\delta(w), x(w), \zeta)=\frac {J_{1}} N \, . 
\end{eqnarray}
Hereafter we shall often abbreviate the counting function 
$Z_1(\delta(w), x(w), \zeta)$ simply by $Z_1(w)$ or 
even by $Z(w)$

\subsection{Difference of the two Bethe quantum numbers $J_1$ and $J_2$ }\label{sec:DTQN}
Let us consider the second Bethe ansatz equation (\ref{eq:BAE22cp}). 
The term corresponding to eq. (\ref{eq:BAELHS1}) is expressed in terms of $a$ in (\ref{eq:a}) 
and $b$ in (\ref{eq:b}) as  
\begin{eqnarray}
\frac{\tan{\lambda_{2}}}{\tanh{\zeta/2}}=a-ib.
\end{eqnarray}
From the real part of the second Bethe ansatz equation (\ref{eq:BAE22cp}) 
we define the second counting function $Z_2(w)$ by   
\begin{eqnarray}
Z_{2}(w) & = {\frac 1 {2 \pi}}  \tan^{-1} \biggl( \frac{a}{1-b} \biggr) + 
 {\frac 1 {2 \pi}} \tan^{-1}\biggl( \frac{a}{1+b}\biggr)   \nonumber \\ 
& \quad +{\frac 1 2} \left(  H(b-1) \mbox{sgn}(a_{-})  + \frac{1}{N} H(\delta) \right). \label{eq:counting2}
\end{eqnarray}
From (\ref{eq:counting1}) and (\ref{eq:counting2}) we derive the difference of 
the two counting functions  as follows.  
\begin{equation}
Z_{2}(w)-Z_{1}(w)= {\frac 1 N} H(\delta) - H(-a) \, .  
\end{equation}
In terms of the Bethe quantum numbers $J_1$ and $J_2$ the difference is expressed as 
$J_2 -J_1 = H(\delta ) - N H(- x) \, $. 
For simplicity, we shift the second Bethe quantum number $J_2$ by $N$, so that we have    
\begin{equation} 
J_2 -J_1 = H(\delta )  \, . \label{eq:J2-J1}
\end{equation}

\subsection{Imaginary part}
We now express $x$ as a function of $w$ through the constraint for the imaginary part of the Bethe ansatz equation (\ref{eq:BAE_2}) to vanish: 
\begin{eqnarray}
\frac{a^2+(b-1)^2}{a^2+(b+1)^2}=\Biggl(\frac{\tanh(\zeta+2\delta)/\tanh(\zeta)-1}
{\tanh(\zeta+2\delta)/\tanh(\zeta)+1}\Biggr)^{\frac{2}{N}} \, . 
\label{eq:abzeta1}
\end{eqnarray}
Let us introduce $X$ by 
\begin{equation}
X \equiv \tan^2{x} ,
\end{equation} 
We express eq. (\ref{eq:abzeta1}) in terms of $X$, $w$ and $t$, as follows.
\begin{eqnarray}
\frac{X(1-w^2 t^2)^2\Bigl/ t^2+\biggl((1-w)+X(-w+w^2 t^2)\biggr)^2}{X(1-w^2 t^2)^2\Bigl/ t^2+\biggl((1+w)+X(w+w^2 t^2)\biggr)^2}\nonumber\\
=\Biggl(\frac{-(1-w)(1-w t^2)}{1+w^2 t^2}\Biggr)^{\frac{2}{N}}\biggl/\Biggl(\frac{(1+w)(1+w t^2)}{1+w^2 t^2}\Biggr)^{\frac{2}{N}} \, . 
\label{eq:abzeta2}
\end{eqnarray}
The equation (\ref{eq:abzeta2})  is equivalent to the quadratic equations 
of $X$ as follows. 
\begin{eqnarray}
A(w)X^2+B(w)X+C(w)=0\label{BAE_ABC}\label{eq:AXBXC}
\end{eqnarray}
where $A(w)$, $B(w)$ and $C(w)$ are defined  by 
\begin{eqnarray}
A(w)&=w^2(1+w t^2)^2\Biggl\{\biggl(\frac{-(1-w)(1-w t^2)}{1+w^{2} t^2}\biggr)^{2}\Biggr\}^{\frac{1}{N}}\\
&-w^2(1-w t^2)^2\Biggl\{\biggl(\frac{(1+w)(1+w t^2)}{1+w^{2} t^2}\biggr)^{2}\Biggr\}^{\frac{1}{N}} \, , \nonumber\\
B(w)&=\biggl\{\frac{(1-w^2 t^2)^2}{ t^2}+2w(1+w)(1+w t^2)\biggr\}\Biggl\{\biggl(\frac{-(1-w)(1-w t^2)}{1+w^{2} t^2}\biggr)^{2}\Biggr\}^{\frac{1}{N}} \, , 
\nonumber\\
&-\biggl\{\frac{(1-w^2 t^2)^2}{ t^2}-2w(1-w)(1-w t^2)\biggr\}\Biggl\{\biggl(\frac{(1+w)(1+w t^2)}{1+w^{2} t^2}\biggr)^{2}\Biggr\}^{\frac{1}{N}}\\
C(w)&=(1+w)^{2}\Biggl\{\biggl(\frac{-(1-w)(1-w t^2)}{1+w^{2} t^2}\biggr)^{2}\Biggr\}^{\frac{1}{N}}-(1-w)^{2}\Biggl\{\biggl(\frac{(1+w)(1+w t^2)}{1+w^{2} t^2}\biggr)^{2}\Biggr\}^{\frac{1}{N}} \, . \nonumber \\
\end{eqnarray}
Here we choose the branch of fractional power function $(1-w)^{2/N}$ as  
\begin{equation} 
(1-w)^{2/N} = \left( (1-w)^2 \right)^{1/N} \, .  \label{eq:branch-choice}
\end{equation}
Thus, $X= \tan^{2}x$ is expressed as a function of $w$ as follows.  
\begin{eqnarray}
X_{\pm} =\frac{1}{2A(w)}\Bigl(-B(w)\pm\sqrt{B(w)^2-4A(w)C(w)}\Bigr) \, . 
\label{eq:tanx}
\end{eqnarray}
We denote $X_+$ and $X_-$ explicitly by $X_p$ and $X_m$, respectively. 
Here we recall that $w$ is a function of $\delta$ through eq. (\ref{eq:def-w}).

In order to determine the regions of $w$ satisfying $X_{\pm} \geq0$, 
we show two lemmas.
\\
\bf{Lemma 1}\\
\rm{} $C(w)\ge0$ and $C(w=1)=0$ for $N\ge2$. \\
\bf{Proof}\\
\rm{} By making use of the fact that $ t = \tanh(\zeta/2) \le 1$ and $0 \le w$, we show  
\begin{eqnarray}
& C(w) \times  (1+ w^2 t^2)^{2/N} \nonumber \\ 
&=(1+w)^{2}\Biggl\{\biggl(-(1-w)(1-w t^{2})\biggr)^{2}\Biggr\}^{\frac{1}{N}}
-(1-w)^{2}\Biggl\{\biggl((1+w)(1+w t^{2})\biggr)^{2}\Biggr\}^{\frac{1}{N}}\nonumber\\
&\geq(1+w)^{2}\biggl\{(1-w)^{2}\biggr\}^{\frac{1}{N}}\biggl\{(1-w)^{2}\biggr\}^{\frac{1}{N}}-(1-w)^{2}\biggl\{(1+w)^{2}\biggr\}^{\frac{1}{N}}\biggl\{(1+w)^{2}\biggr\}^{\frac{1}{N}}\nonumber\\
&\left\{ \begin{array}{ll}
    =0 & (N=2) \\
    \geq(1+w)^{2}(1-w)^{2}-(1-w)^{2}(1+w)^{2}=0 & (N>2).
  \end{array} \right.
\end{eqnarray} 
\bf{Lemma 2}\\
\rm{}If $A(w)\geq0$, then $B(w)\geq0$\ \ $(N\ge2)$\\
\bf{Proof}\\
\rm{} We express the inequality: $A(w)\geq0$ for $0 < w < 1/t$ as follows. 
\begin{eqnarray}
\biggl\{\Bigl(\frac{-(1-w)(1-w t^{2})} {(1+w) (1 + w t^{2})} \Bigr)^{2}
\biggr\}^{\frac{1}{N}} 
&\geq \frac { (1- w t^2)^{2}} {(1+ w t^2)^{2}}.  
\end{eqnarray}
\rm{}By making use of  this inequality, we have  
\begin{eqnarray}
B(w) \geq&\Biggl[\biggl\{\frac{(1-w^{2} t^{2})^{2}}{ t^2}+2w(1+w)(1+w t^{2})\biggr\}\Bigl(\frac{1-w t^{2}}{1+w t^{2}}\Bigr)^{2}\nonumber\\
&-\biggl\{\frac{(1-w^{2} t^{2})^{2}}{ t^2}-2w(1-w)(1-w t^{2})\biggr\}\Biggr]\Biggl\{\Bigl(\frac{(1+w)(1+w t^{2})}{1+w^{2} t^{2}}\Bigr)^{2}\Biggr\}^{\frac{1}{N}}\nonumber\\
\label{eq:Bineq}
\end{eqnarray}
Since we have  $\Biggl\{\Bigl(\frac{(1+w)(1+w t^{2})}{1+w^{2} t^{2}}\Bigr)^{2}\Biggr\}^{\frac{1}{N}}\geq0$, we  evaluate 
\begin{eqnarray}
& B(w)/ \Biggl\{\Bigl(\frac{(1+w)(1+w t^{2})}{1+w^{2} t^{2}}\Bigr)^{2}\Biggr\}^{\frac{1}{N}} 
\nonumber \\ 
& \geq  \frac{(1-w^{2} t^{2})^{2}}{ t^2}\Bigl(\frac{1-w t^{2}}{1+w t^{2}}\Bigr)^{2}+2w(1+w)(1+w t^{2})\Bigl(\frac{1-w t^{2}}{1+w t^{2}}\Bigr)^{2}\nonumber\\
&+2w(1-w)(1-w t^{2})-\frac{(1-w^{2} t^{2})^{2}}{ t^2}\nonumber\\
&=\frac{2w}{(1+w t^{2})^{2}}2(1-w^{2} t^{2})w^{2} t^{2}(1- t^{2})\nonumber\\
&\geq0 \, . 
\end{eqnarray}
We have thus shown that if $A(w)\geq0$ then $B(w)\geq0$.

\subsection{Conditions of $X=\tan^2 x$ being positive as a function of $w$}
\label{sec:CX}

Let us first consider the case of $w\neq 1$. 
We recall Lemma 1 that we have $C \ge 0$. 
In the case of  $X_p$ we  have four cases:  
(i) $A > 0$ and $B>0$;   (ii) $A > 0$ and $B<0$;   
(iii) $A < 0$ and $B>0$;   (iv) $A < 0$ and $B<0$. 
In each case we check whether $X_{p}$ is positive:
\begin{description}
 \item[($\rm{i}$)]$A>0,B>0$: 
$X_p=\frac{1}{2A}\biggl(-|B|+\sqrt{|B|^{2}-4|A||C|}\biggr)<0 $; 
 \item[($\rm{i\hspace{-.1em}i}$)]$A>0, B<0$:
$X_p=\frac{1}{2A}\biggl(|B|+\sqrt{|B|^{2}-4|A||C|}\biggr)>0$; 
 \item[($\rm{i\hspace{-.1em}i\hspace{-.1em}i}$)] $A<0, B>0$:
$X_p=\frac{-1}{2A}\biggl(|B|-\sqrt{|B|^{2}+4|A||C|}\biggr)<0$;  
 \item[($\rm{i\hspace{-.1em}v}$)]$A<0,B<0$: 
$X_p=\frac{-1}{2A}\biggl(-|B| - \sqrt{|B|^{2}+4|A||C|}\biggr)<0$.  
\end{description}
We have  $X_{p} > 0$ only for ($\rm{i\hspace{-.1em}i}$). 
However, it is not allowed due to Lemma 2.

For  $X_{m}$ thanks to the fact that $C\geq0$, we 
consider only the signs of $A$ and $B$. 
In each case of $A, B$, we check whether $X_{m}(w)$ is positive.
\begin{description}
 \item[($\rm{v}$)]$A>0, B>0$: 
$X_m=\frac{1}{2A}\biggl(-|B|-\sqrt{|B|^{2}-4|A||C|}\biggr)<0$;   
 \item[($\rm{v\hspace{-.1em}i}$)]$A>0, B<0$: 
$X_m=\frac{1}{2A}\biggl(|B|-\sqrt{|B|^{2}-4|A||C|}\biggr)>0$; 
 \item[($\rm{v\hspace{-.1em}i\hspace{-.1em}i})$]$A<0,B>0$: 
$X_m=\frac{-1}{2A}\biggl(|B|+\sqrt{|B|^{2}+4|A||C|}\biggr)>0$; 
 \item[($\rm{v\hspace{-.1em}i\hspace{-.1em}i\hspace{-.1em}i}$)]$A<0, B<0$:
$X_m=\frac{-1}{2A}\biggl(-|B|-\sqrt{|B|^{2}+4|A||C|}\biggr)>0$ .
\end{description}
We have $X_{m} > 0$ for ($\rm{v\hspace{-.1em}i}$), ($\rm{v\hspace{-.1em}i\hspace{-.1em}i})$ and 
($\rm{v\hspace{-.1em}i\hspace{-.1em}i\hspace{-.1em}i}$).  
But ($\rm{v\hspace{-.1em}i}$) is not allowed due to Lemma 2. 

We now consider the case of $w=1$. Since $C(1)=0$, 
eq. (\ref{eq:AXBXC}) is given by 
\begin{eqnarray}
A(w=1)X^{2}+B(w=1)X=0.
\end{eqnarray}
Since $A(1) < 0$ and $B(1) < 0$,  we have $X_{m}=0$. 
We thus have $x=0$ and $\delta=0$ ($w=1$). Here we remark that 
it gives a singular solution shown in section \ref{sec:SS}.

We therefore conclude the following. 
\\
\bf{Proposition 1}\\
\rm{} $X_{\pm}(w) >0$ if and only if we have $A(w) < 0$ and for $X_{-}(w)$.  

Hereafter we denote $X_m(w)=X_{-}(w)$ simply by $X(w)$. 

In summary,  for a given  Bethe quantum number  $J_{1}$, 
we evaluate $\delta$ and $x$  numerically 
by evaluating $w$ with the counting function $Z(w)$ in eq. (\ref{eq:counting1})
and then center $x$ by making use of (\ref{eq:tanx}) for $X_{-} (= X_m)$.  
Here we remark that the second Bethe quantum number $J_{2}$ is given by eq. (\ref{eq:J2-J1}).

\section{Monotonicity of the counting function}\label{sec:monotone}
\label{sec:monotone}

\subsection{Domains of $w$ for narrow pairs and wide pairs: 
$A(w)$ has at most one zero in $0<w<1$ and exactly one zero in $1 < w < 1/t$ }
\label{sec:A0}

Let us recall that the center $x$ and the deviation $\delta$ of a two-string solution 
$\lambda_1$ and $\lambda_2$ have been defined in eqs.  (\ref{2string}).  
We call two-string solutions with  $\delta < 0$ {\it narrow pairs}  
and those of $\delta > 0$ {\it wide pairs}.  
The intervals of $w$ satisfying $0<w<1$ and $1 < w < 1/t$ correspond to 
two-string solutions  (\ref{2string})  with $\delta < 0$ and those of $\delta > 0$, respectively, through eq. (\ref{eq:def-w}).

We shall show that  the equation: $A(w)=0$ has at most one solution in the interval $0<w<1$. 
We shall denote it by $w_1$. If there is no solution to it, we show that $A(w) < 0$ for 
 $0<w<1$ and $A(0)=0$, and define $w_1$ by $w_1=0$. 
We shall also show that  the equation: $A(w)=0$ has one and only one solution 
in the interval $1<w<1/t$. We denote it by $w_3$. 

We remark that due to Proposition 1 of section \ref{sec:CX} 
it  is fundamental to consider the regions of $w$ where 
we have $A(w) <0$.

\subsubsection{Derivation of the stable regime criterion 
from  the derivative of $A(w)$ at $w=0$}
Let us define $\widetilde{ A}(w)$ by 
\begin{equation} 
\widetilde{A}(w) = - A(w) \left( \frac {(1 + w^2 t^2)} {(1+ w t^2)(1-wt^2)} \right)^{2/N}/w^2   
\end{equation} 
Explicitly we have 
\begin{equation} 
\widetilde{A}(w) 
= \left( (1-wt^2)^{N-1} (1+w) \right)^{2/N} - 
\left( (1 + wt^2)^{N-1} (1-w) \right)^{2/N} \,. 
\end{equation}
It is clear that for $w$ satisfying $0 < w <1$  
the equation $A(w)=0$ is equivalent to the equation $\tilde{A}(w)=0$,   
since each of the factors $1 + w^2 t^2$, $1+ w t^2$ and $1-wt^2$ 
does not vanish for $0 < w <1$.
We shall present the derivative of $\widetilde{A}(w)$ in section \ref{sec:derivatives}.  

The derivative of $\widetilde{ A}(w)$ at $w=0$ is thus given by  
\begin{equation} 
{\frac {d \widetilde{ A}} {d w}}(w=0)  = 4 t^2 \left( \frac {N_{\zeta}} N -1  \right) 
\end{equation}
Here we recall that $N_{\zeta}$ has been defined in eq. (\ref{eq:Nzeta}). 
The derivative $d\widetilde{ A}/dw$ at $w=0$ is negative if 
$N > N_{\zeta}$, while it is positive  otherwise.  
The criterion whether $N > N_{\zeta}$ or not, i.e.  
 $N$ and $\Delta$ are in the stable regime or not,  plays a fundamental role in this paper. 
Here we recall that $t=\tanh(\zeta/2)$, and also  
that the XXZ anisotropy $\Delta$ is given  by $\Delta= \cosh \zeta$.

\subsubsection{Stable regime for $0 < w< 1$} 

When $(N-1) t^2 > 1$ holds, the derivative ${{d \widetilde{ A}}/{d w}}$ is negative  at $w=0$.    
We thus have 
\begin{eqnarray} 
& {\frac {d \widetilde{ A}} {d w}}(w)  = 2 t^2  
\left( \frac {1 + w t^2 } {1-w}  \right)^{1-{\frac 2 N}}  \times 
\nonumber \\ 
& \, \times 
\Bigg[ 
 \left( \frac {N_{\zeta}} N -1 + w   \right)   
 -  \left( 1-  \frac {N_{\zeta}} N + w \right) 
 \left( \frac {(1-w) (1 - w t^2) } {(1+w) (1 + wt^2) }  \right)^{1-{\frac 2 N}}   \Bigg] \,. 
\label{eq:dAt}
\end{eqnarray}
The derivative (\ref{eq:dAt}) is negative at $w=0$, 
while it becomes positive as $w$ increases for $0 < w < 1$.  
In the second term of eq. (\ref{eq:dAt}) it is easy to show that the following 
inequality holds  for $0 < w < 1$. 
\begin{equation}
\left| \left( \frac {(1-w) (1 - w t^2) } {(1+w) (1 + wt^2) }  \right)^{1-{\frac 2 N}} \right| < 1.   
\end{equation}
The dominant term in eq. (\ref{eq:dAt}), 
$\left( N_{\zeta}/N  - 1 + w \right) $, becomes positive when 
$w$ is larger than $1 - N_{\zeta}/N$, and hence 
there exists a point $w_a$ such that 
the derivative becomes positive for any $w$ satisfying $w_a < w < 1$.

We note that  $\widetilde{A}=0$ and $d \widetilde{A}/dw<0$ at $w=0$ .   
Therefore,  $\widetilde{A}(w)$ decreases at $w=0$ and is negative for $w > 0$ 
at least in the neighborhood of $w=0$.  However,  it turns to increase  
since $d \widetilde{A}/dw > 0$ for $w> w_a$. 
Moreover, $\widetilde{A}(w)$ is positive at $w=1$ 
\begin{equation}
\widetilde{A}(1)= 2^{2/N} (1-t^2)^{2-2/N} > 0 \, . \label{eq:Aat1}
\end{equation} 
Therefore,  $\widetilde{A}(w)$ vanishes only at one point in     
$0 < w < 1$.  We denote it by $w_1$.

\subsubsection{Unstable regime for $0 < w < 1$} 

In the unstable regime we have $N < N_{\zeta}$, and the derivative  (\ref{eq:dAt})  at $w=0$ is 
positive. The dominant term of (\ref{eq:dAt})  
increases and is always positive for $w$ satisfying $0 < w < 1$. 
Therefore, the derivative of $\widetilde{A }$ is always positive for 
$0 < w < 1$.  Here we remark that $\widetilde{A}(0) =0$.   
It follows that $\widetilde{A}(w) > 0 $ for $w$ satisfying $0 < w< 1$.

\subsubsection{Uniqueness of the zero for $A(w)$ in the interval $1 < w < 1/t$ } 

We recall that for the fractional power  $(1-w)^{2/N}$ we have chosen the branch:  
$(1-w)^{2/N}= \left( (1-w)^2 \right)^{1/N}$. 
It follows that  $\widetilde{A}(w)$ for $w > 1$ is given by 
\begin{equation} 
\widetilde{A}(w) 
= \left( (1-wt^2)^{N-1} (1+w) \right)^{2/N} - 
\left( (1 + wt^2)^{N-1} (w-1) \right)^{2/N} \,. 
\end{equation}
We thus have for $w > 1$  
\begin{eqnarray} 
& {\frac {d \widetilde{ A}} {d w}}(w)  = 2 t^2 
\left( \frac  {1 + w t^2} {w-1}  \right)^{1-{\frac 2 N}}  \Bigg[
- \left( w -1 + \frac {N_{\zeta}} N   \right)    
\nonumber \\ 
& \quad - \left( w + 1 - \frac {N_{\zeta}} N  \right) 
 \left( \frac {(w-1) (1 - w t^2) } {(1+w) (1 + wt^2) }  \right)^{1-{\frac 2 N}}   \Bigg] \,. 
\label{eq:dAt2}
\end{eqnarray}
The term: $w - 1 + N_{\zeta}/N$ is always positive for $w>1$.   It leads to 
the dominant term in  (\ref{eq:dAt2}), and 
hence the derivative (\ref{eq:dAt2}) is always negative in the interval  $1 < w < 1/t $ 
for both the stable and unstable regimes. 
It follows that $\widetilde{A}(w)$ is monotonically decreasing in the interval 
$1 < w < 1/t$.   

We have shown that $\widetilde{A}(1)>0 $ in eq. (\ref{eq:Aat1}). 
It is eay to show that $\widetilde{A}(w)$ is negative at $w=1/t$.  
\begin{equation} 
\widetilde{ A} (1/t) = t^{-2/N} (1+t)^2 \left( \frac {1-t}{1+t} \right)^{2/N} 
\left\{ \left( \frac {1-t}{1+t} \right)^{2 - 4/N} -1 \right\} < 0 \,.  
\end{equation} 
Therefore, $\widetilde{ A}(w)$ vanishes only at one point of $w$ satisfying $1 < w < 1/t$. 
We denote it by $w_3$. 
Thus, the equation $A(w)=0$ has the unique solution $w=w_3$ in the interval $1 < w < 1/t$.

\subsection{Approximate values of zeros $w_1$ and $w_3$}\label{sec:w1w3} 

\subsubsection{In the interval $0 < w < 1$ of narrow pairs} 

The equation $A(w)=0$ is equivalent to 
\begin{equation}
\left( \frac {1 -wt^2} {1+wt^2} \right)^{N-1} = \frac {1 - w} {1 +w} \, . \label{eq:Aw1}
\end{equation}
If $N$ is very large,  the left hand side of eq. (\ref{eq:Aw1}) 
is approximately expressed with the exponential function. 
\begin{equation}
\left( \frac {1 -wt^2} {1+wt^2} \right)^{N-1} \approx \exp \left( -2 (N-1) t^2 w \right) \,. 
\label{eq:A-LHS}
\end{equation}
We express eq. (\ref{eq:Aw1}) in terms of the hyperbolic tangent function 
\begin{equation}
w = \tanh\left( (N-1) t^2 w \right) \,.   \label{eq:MFA}
\end{equation}

Let us solve eq. (\ref{eq:MFA}) approximately. 
We consider two cases: (i) If $(N-1) t^2 < 1$ holds, then eq. (\ref{eq:MFA})
has no nonzero solution. Therefore, in the unstable regime we have $w_1=0$;  
(ii) If $(N-1) t^2 > 1$ holds,  eq. (\ref{eq:MFA})
has a nonzero solution. Suppose that $(N-1) t^2$ is much larger than 1. 
We then approximate $w_1$ by 
\begin{equation}
w_1 = 1 - 2 \exp\left(-2 (N-1)t^2 \right) + O(\exp\left(- 4 (N-1)t^2 \right) \, . 
\end{equation}
We have thus shown approximately that  $w_1$ approaches 1 exponentially as $N$ increases. 

\subsubsection{In the interval $1 < w < 1/t$ of wide pairs} 

We can show that $w_3$ satisfies 
\begin{equation}
\left( \frac {1 -wt^2} {1+wt^2} \right)^{N-1} = \frac {w- 1} {w+ 1} \, . \label{eq:Aw2}
\end{equation}
We then approximate $w_3$ by 
\begin{equation}
w_3 = 1 + 2 \exp\left(-2 (N-1) t^2 \right) + O( \exp\left(-4 (N-1) t^2 \right))
\end{equation}
We have thus shown approximately that  $w_3$ approaches 1 exponentially as $N$ increases. 

We remark that in section \ref{sec:RSH} we shall rigorously  show that $w_1$ and $w_3$ approach
1 exponentially fast with respect to $N$ as the number of sites $N$ increases.

\subsection{Graphs of $B$ and $C$ versus $w$}
\subsubsection{Monotonicity of $\widehat B(w)$}
Let us define $\widehat{ B}(w)$ by 
\begin{equation} 
\widehat{B}(w) =  B(w) \left( \frac {(1 + w^2 t^2)} {(1+ w t^2)(1-wt^2)} \right)^{2/N}/w   
\, . 
\end{equation} 
Explicitly we have 
\begin{eqnarray}
\widehat{B}(w) &=\biggl\{\frac{(1-w^2 t^2)^2}{w t^2 }+2(1+w)(1+w t^2)
\biggr\} \biggl (\frac{1-w}{1+w t^2} \biggr)^{2/N} \nonumber\\
&- \Biggl\{\frac{(1-w^2 t^2)^2}{w t^2}-2(1-w)(1-w t^2) 
\Biggr\} \biggl(\frac {1+w}{1 - w t^2} \biggr)^{2/N} \, . 
\end{eqnarray}
For $0 < w < 1$ we expand $\widehat{ B}(w)$ 
with respect to $w$ through formula (\ref{eq:exp}) 
\begin{eqnarray} 
\widehat{ B}(w) & = 4 \left(1 - \frac {N_{\zeta}} N \right) \nonumber \\  
& + \Bigg\{ - {\frac 4 3}\frac {N_{\zeta}} N + 4 t^2 \Bigg(1 - {\frac {2 (1+2 t^2)} 3} 
\frac {N_{\zeta} } {N} + (1+t^2)   \frac {N_{\zeta}^2 } {N^2}  -  {\frac {2 t^2}  3}  {\frac {N_{\zeta}^2 } {N^2} }  \Bigg) \Bigg\} w^2 \nonumber \\ 
& +  O(w^4) \,  . \label{eq:Bhat-w2}
\end{eqnarray} 
We have 
\begin{equation} 
\widehat{B}(0)= 4 \left(1 - \frac {N_{\zeta}} N  \right)  
 \, . 
\end{equation}
Hence,  $\widehat{B}(0)$ is positive in the stable regime, 
while negative in the unstable regime.  
It is easy to show that $\widehat{ B}(w)$ is negative at $w=w_2 =1$. 
\begin{equation}
\widehat{B}(1) = - \frac 4 {t^2} \left( \frac {1-t^2} 2\right)^{2-2/N} < 0 \,. 
\end{equation} 
Similarly we can show 
\begin{equation}
\widehat{B}(1/t) =  \frac {2 (1+t)^2} {t^{1+2/N}} \left( \frac {1-t} {1+t}  \right)^{2/N} 
\left\{ 1- \left(  {\frac {1-t} {1+t} }   \right) ^{2-4/N} \right\}  > 0 \,. 
\end{equation} 

The derivative of $\widehat{ B}(w)$ with respect to $w$ is presented explicitly in Appendix B. 
When $N$ is large, $w_1$ is close to 1 in the stable regime and 
$w_3$ is close to 1 both in the stable and unstable regimes, as shown in section \ref{sec:w1w3}. 
It is clear that  terms with factor $|w_2-w|^{-1}$  become dominant for large $N$  if the range of 
$w$ is limited in the neighborhood of 1 since $w_1$ or $w_3$ is close to 1. 
In the interval $w_1 < w < w_2$ of narrow pairs  we show that 
the derivative of $\widehat{ B}(w)$  is negative  for large $N$, 
since all terms with the factor $|w_2-w|^{-1}$ give only negative contributions, 
as shown in eq. (\ref{eq:dBhat}).   
In the interval $w_2 < w < w_3$ of wide pairs 
we show that the derivative of $\widehat{ B}(w)$ is positive  for large $N$, 
since all terms with the factor $|w_2-w|^{-1}$ give only 
positive contributions.

In the interval $w_1 < w < w_2$ of narrow pairs, 
in the stable regime when $N$ is large  we prove that ${\widehat B}(w)$ is monotonically  decreasing with respect to $w$ since $d{\widehat B}(w)/dw$ (\ref{eq:dBhat})
is negative for  $w_1 < w < w_2$. 
In the unstable regime, however,  we conjecture that ${\widehat B}(w)$ is  monotonically  decreasing with respect to $w$.    We confirm it  at least near $w=0$ by 
calculating that the coefficient of the second power $w^2$ 
in (\ref{eq:Bhat-w2}) is negative.  
In fact, the dominant term in the coefficient $-4 N_{\zeta}/ 3N + 4t^2$ is negative 
if $N <  N_{\zeta}/(3 t^2)$, which holds  for small $t$ in the unstable regime.

In the interval $w_2 < w < w_3$ of wide pairs, when $N$ is large  
both in the stable and unstable regimes it follows  
 that ${\widehat B}(w)$ is  monotonically increasing with respect to $w$, since   
$d{\widehat B}(w)/dw$ is positive in the whole interval 
of $w$ satisfying $w_2 < w < w_3$.

\subsubsection{Monotonicity of $\widehat C(w)$}

Let us define $\widehat{ C}(w)$ by 
\begin{equation} 
\widehat{C}(w) =  C(w) \left( \frac {(1 + w^2 t^2)} {(1+ w t^2)(1-wt^2)} \right)^{2/N}/w   
\end{equation} 
Explicitly we have 
\begin{eqnarray}
\widehat{C}(w)&={\frac {(1+w)^{2} } w} \biggl(\frac{1-w}{1+w t^2} \biggr)^{2/N} 
-{\frac {(1-w)^{2}} w}  \biggl(\frac{1 + w}{1 - w t^2}\biggr)^{2/N} \, .  
\end{eqnarray}
For $0 < w < 1$ we expand $\widehat{ C}(w)$ 
with respect to $w$ through formula (\ref{eq:exp}) 
\begin{eqnarray} 
\widehat{ C}(w) & = 4 \left(1 - t^2 \frac {N_{\zeta}} N \right) 
 + \Bigg\{ - {\frac 4 3} t^2 \left( (2-t^2)^2 +3\right)  \frac {N_{\zeta}} N 
\nonumber \\   
& + 4 t^4  (3- t^2)  \frac {N_{\zeta}^2 } {N^2}   - 
 {\frac {8 t^6}  3}  {\frac {N_{\zeta}^3 } {N^3} }  \Bigg) \Bigg\} w^2  +  O(w^4)  \,  . 
\label{eq:Chat-w2}
\end{eqnarray}

It is easy to show that 
\begin{equation}
\widehat{C}(0) = 4 \left(1 - t^2 \frac {N_{\zeta}} N  \right) \, .  
\end{equation} 
It is positive at $w=0$ since we have $N > 2 > 1+t^2 = t^2 N_{\zeta}$. 
Straightforwardly we calculate 
\begin{equation}
\widehat{C}(1/t) = \frac {(1+t)^2} {t^{1+2/N}} \left( \frac {1-t} {1+t}  \right)^{2/N} 
\left\{ 1 - \left( \frac {1-t} {1+t}  \right)^{2-4/N} \right\} > 0 
\, .  
\end{equation} 

We can argue the monotonicity of $\widehat{ C}(w)$ 
similarly as $\widehat{ B}(w)$.  
The derivative of $\widehat{ C}(w)$ with respect to $w$ is given  in Appendix C. 
We recall that when $N$ is large $w_1$ is close to 1 in the stable regime 
and $w_3$ is close to 1 in the both stable and unstable regimes,  
and also  that terms with factor $|w_2-w|^{-1}$  become 
dominant when the range of $w$ is restricted in the neighborhood of $w_2=1$. 
Therefore, in the interval $w_1 < w < w_2$ of narrow pairs  we show that 
the derivative of $\widehat{ C}(w)$  is negative  for large $N$, 
since all terms with the factor $|w_2-w|^{-1}$ in eq. (\ref{eq:dChat}) 
give only negative contributions;  
In the interval $w_2 < w < w_3$ of wide pairs 
we show that the derivative of $\widehat{ C}(w)$ is positive  for large $N$, 
since all terms with the factor $|w_2-w|^{-1}$ give only 
positive contributions.

In the interval $w_1 < w < w_2$ of narrow pairs, 
when $N$ is large in the stable regime 
we show that ${\widehat C}(w)$ is  monotonically  decreasing with respect to $w$.  
since  $d{\widehat C}(w)/dw$ is negative throughout the interval   $w_1 < w < w_2$. 
However, in the unstable regime 
at least near $w=0$ we show that it is monotonically decreasing 
by showing that the coefficient of the second power $w^2$ in eq. (\ref{eq:Chat-w2}) is negative.

In the interval $w_2 < w < w_3$ of wide pairs, we show that 
${\widehat C}(w)$ is  monotonically increasing with respect to $w$ in the both 
stable and unstable regimes.    We  prove it  when $N$ is large 
since $d{\widehat C}(w)/dw$ is positive for  $w_2 < w < w_3$.

\subsection{Monotonicity of $X$}

We show the monotonicity of $X(w)$ as a function of $w$ 
by making use of the following formula:
\begin{eqnarray}
& \frac {d X} {d w}(w) =  {\frac 1 {2 {\widehat A}} }  {\frac {d {\widehat B}}  {d w} }  
\frac  {\widehat{ B} + \sqrt{ {\widehat B}^2 + 4 {\widehat A} {\widehat C} } }
	 {\sqrt{ {\widehat B}^2 + 4 {\widehat A} {\widehat C} } } \nonumber \\ 
& \quad + {\frac 1 { {\widehat A} \sqrt{ {\widehat B}^2 + 4 {\widehat A} {\widehat C}  } } }
\left\{  {\widehat A} \frac {d {\widehat C} } {dw} 
- {\frac 1 {2 {\widehat A} } }  {\frac {d {\widehat A} }  {d w} } 
\left( {\widehat B}^2 + {\widehat B} 
{\sqrt{ {\widehat B}^2 + 4 {\widehat A} {\widehat C} }} + 2 {\widehat A}  {\widehat C} 
\right) \right\} 
 \, .  
\nonumber \\  
\label{eq:XmABC}
\end{eqnarray}
Here we have defined $ {\widehat A}(w)$ by 
\begin{equation} 
 {\widehat A}(w) = w {\widetilde A}(w) \,. 
\end{equation}
We shall also make use of the following lemma 
\par \noindent 
{\bf Lemma 3}\\
\rm{} $\beta^2 - \beta \sqrt{\beta^2 +  2 \alpha} + \alpha > 0$ for $\alpha, \beta > 0$. \\

It is easy to show 
\begin{equation}
\frac {d \widehat{ A}} {d w} = {\widetilde{A}} + w \frac {d \widetilde{A}} {dw}  
\, . 
\end{equation}
Since $d {\widetilde A}(w)/dw> 0$  and ${\widetilde A}(w)  > 0$ for $w_1 < w< w_2$,  
it follows that $d {\widehat A}(w)/dw> 0$ and ${\widehat A}(w) > 0$  for $w_1 < w< w_2$.  In the interval $w_2 < w < w_3 $ of wide pairs, in the both stable and unstable regimes  we show when $N$ is large  that the derivative of  ${\widehat{A}}(w)$ is negative, since all terms with the factor $|w_2-w|^{-1}$  give only negative contributions.

In the interval $w_2 < w < w_3 $ of wide pairs, 
in the both stable and unstable  regimes for large $N$, 
the derivative of $\widehat{A}(w)$ is negative while  
the derivatives of $\widehat{B}(w)$ and  $\widehat{C}(w)$ are positive. 
It follows from (\ref{eq:XmABC}) and Lemma 3 that 
${d X}/{d w}(w)$  is positive.   
Hence, $X(w)$ is monotone increasing with respect to $w$ in the interval of $w$ with $w_2 < w < w_3$.

In the interval $w_1 < w < w_2 $ of narrow pairs, 
 in the stable regime for large $N$ we can show that the derivatives of $\widehat{  B}$ and $\widehat{ C}$ are negative. Here we recall that the positivity of  the derivative of ${\widehat  A}$ has already been shown for any $N$.  
It follows from (\ref{eq:XmABC}) with Lemma 3 that 
 ${d X}/{d w}(w)$  is negative. 
Hence, $X$ is monotone decreasing with respect to $w$ for $w_1 < w < w_2 $.

In the interval $w_1 < w < w_2 $ of narrow pairs, 
 in the unstable regime  
 we have shown  the negativity of the derivatives of $\widehat{  B}$ and $\widehat{ C}$ 
only near $w=0$.  We have a conjecture that  
the derivatives of $\widehat{  B}$ and $\widehat{ C}$ 
are negative throughout the whole interval of $w$ with $w_1 < w < w_2 $. 
If we assume the conjecture,  it follows from (\ref{eq:XmABC}) with Lemma 3 that 
 ${d X}/{d w}(w)$  is negative. and hence $X$ is monotonically decreasing 
with respect to $w$ for $w_1 < w < w_2 $.

\subsection{Monotonicity of the counting function $Z(w)$}

\subsubsection{Analytical argument}

Let us recall the definition of the counting function 
(\ref{eq:couting1}):
\begin{eqnarray}
Z(w) & = \frac 1 {2 \pi} \tan^{-1} \left( \frac a {1-b} \right) 
+ \frac 1 {2 \pi} \tan^{-1} \left( \frac a {1+ b} \right) 
\nonumber \\ 
& + \frac 1 2 \left( H(b-1) + 2 H(1-b)H(-a) - {\frac 1 N}H(\delta)  \right) \,  . 
\label{eq:CF1}
\end{eqnarray}
Let us define $\Delta Z(w)$ by 
\begin{eqnarray}
\Delta Z(w) = Z(w) - \frac 1 2 \left( H(b-1) + 2 H(1-b)H(-a) - {\frac 1 N}H(\delta)  \right) 
 \end{eqnarray} 
We can show \cite{DI2018}
\begin{eqnarray}
& \tan\left( - 2 \pi \Delta Z(w) \right)   = \frac {2a}  {a^2+b^2-1} 
\nonumber \\ 
= & \quad \frac {2 \sqrt{X} (1 +X w^2 t^2) (1-w^2 t^2)/t} 
{X^2 w^2 (1- w^2 t^4) + X ( t^{-2} - 2 w^2 t^2 + w^4 t^2 ) + (w^2-1) }    
\label{eq:abc}
\end{eqnarray}

Let us now argue that the counting function $Z(w)$  monotonically increases  for wide pairs 
(i.e., for $w$ satisfying $w_2 < w < w_3$) and decreases for narrow pairs (i.e., for $w$ satisfying $w_1 < w < w_2$).

For wide pairs ($w_2 < w < w_3$), the range of parameter $w$ is exponentially narrow with respect to $N$ 
if the number of sites $N$ is very large: $w_3 - w_2= O (\exp(-N/2))$. 
We therefore assume that the expression (\ref{eq:abc}) can be well approximated 
by setting $w=1$. At $w=1$ we have 
\begin{equation}
\left. \frac {2a} {a^2 + b^2 -1} \right|_{w=1} = \frac {2(1-t^2)t} {\sqrt{X}} \,. 
\label{eq:Xwt}
\end{equation} 
Here we recall that we have shown that $X$ increases monotonically 
with respect to $w$ in the interval  $w_2 < w < w_3$. 
It is clear that $\tan\left( - 2 \pi \Delta Z(w) \right)$ decreases monotonically 
with respect to $w$. 
It thus follows that the counting function $Z(w)$ increase monotonically 
with respect to $w$ for wide pairs, i.e. for $w$ satisfying $w_2 < w< w_3$.

For narrow pairs ($w_1 < w < w_2$) in the stable regime,   
the range of parameter $w$ becomes exponentially narrow with respect to $N$ 
if the number of sites $N$ is very large: $w_2 - w_1= O (\exp(-N/2))$. 
We therefore assume that the expression (\ref{eq:abc}) can be well approximated 
by setting $w=1$. At $w=1$ we have the same expression (\ref{eq:Xwt}) for 
$\tan\left( - 2 \pi \Delta Z(w) \right)$.  
Here we recall that we have shown that $X$ decreases monotonically 
with respect to $w$ in the interval  $w_1 < w < w_2$ if the range 
of $w$ such as $|w_1-1|$ is very small.  
It is clear from eq. (\ref{eq:Xwt}) 
that $\tan\left( - 2 \pi \Delta Z(w) \right)$ decreases monotonically 
with respect to $w$. 
It  thus follows  that the counting function $Z(w)$ increases monotonically 
with respect to $w$ for narrow pairs, i.e. for $w$ satisfying $w_1 < w< w_2$ 
if $N$ and $\Delta$ is in the stable regime and $N$ is very large.

\begin{figure}[htbp]
\begin{center}
 \includegraphics[clip,width=10cm]{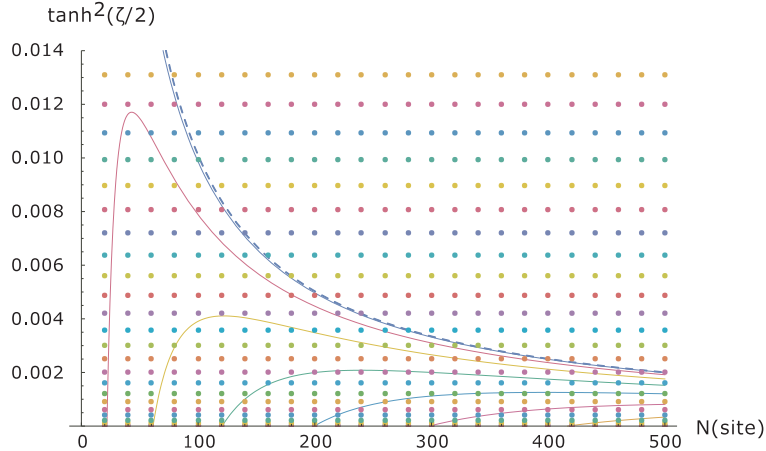}
\end{center}
 \caption{Points where the monotonicity of the counting function 
$Z(w)$ has been checked numerically.  }
    \label{fig:test} 
\end{figure}

\subsubsection{Numerical confirmation}
We have checked the monotonicity of the counting function $Z(w)$ directly by plotting the graphs of $Z(w)$ versus $w$ for many different values of 
$\Delta$ and $N$ in the both stable and unstable regimes. 
The points are plotted together with the boundary line between the stable and unstable regimes and the collapse solution regions. 
in Fig. \ref{fig:test}.

%
%
\section{Counting complex solutions analytically }\label{sec:CCS}

\subsection{Two domains of definition in the counting function for narrow and wide pairs}

Let us recall that due to Proposition 1 of section \ref{sec:CX} 
we consider the regions of $w$ where $A(w) <0$ holds. 
We define the domain of definition for the counting function $Z(w)$   
by the set of values of $w$ where $X(w)$ is positive 
and the counting function $Z(w)$ is defined. 
Furthermore, we have shown in section \ref{sec:A0} that the number of zero of  $A(w)$ for $w$ satisfying $0<w<1$ is at most one, denoted by $w_1$, 
and also that $A(w)=0$ has a zero for $w$ satisfying $1 < w <1/t$, denoted by $w_3$.  

\begin{figure}[htbp]
\begin{center}
 \includegraphics[clip,width=8.0cm]{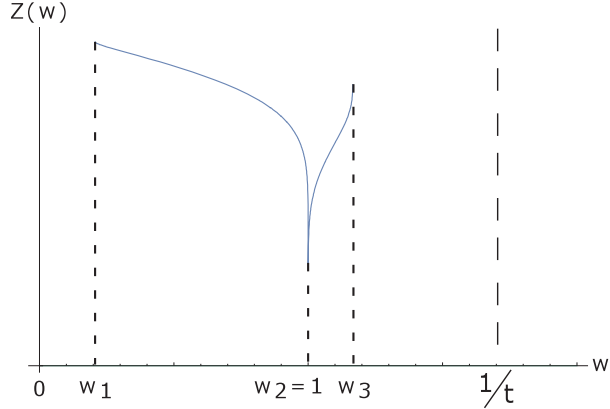}
\end{center}
\caption{Graph of the counting function $Z(w)$ versus variable $w$. 
We assign the counting number in the vertical axis and $w$ in the horizontal axis.}
\label{fig:count}
\end{figure}

It follows that 
the domain of definition for the counting function $Z(w)$ 
is given by the interval between $\inf\{w|A(w)\leq0\}$ and 
$\sup\{w|A(w)\leq0\}$,  since $A(w=1)<0$ and $A(w)$ is continuous. 
We recall that we have defined $w_{1}, w_{2}, w_{3}$ as follows.
\begin{eqnarray}
w_{1}&\equiv \inf\{w|A(w)\leq0\}\\
w_{2}&\equiv 1\\
w_{3}&\equiv \sup\{w|A(w)\leq0\}
\end{eqnarray}
We obtain the counting function, as shown in figure \ref{fig:count}.

The monotonicity of $Z(w)$ has been shown analytically in section 3  if $N$ is very large in the stable regime for $w$ with $w_1 < w < w_3$ 
(both narrow and wide pairs)
and in the unstable regime   for $w$ with $w_2 < w < w_3$ (wide pairs).  
Furthermore, the monotonicity of $Z(w)$ has been numerically confirmed  
at many points in the diagram of $\tanh^2(\zeta/2)$ versus $w$,  as shown in figure 
\ref{fig:test}. 

We thus assume in section 4 that in the case of $x>0$ (or $\tan x > 0$) the counting function $Z(w)$ is monotonically decreasing in the interval of $w$ satisfying 
$w_{1}< w <w_2$ and $Z(w)$ is monotonically increasing in the interval of $w$ satisfying $w_2 < w < w_{3}$,  
and also that in the case of $x<0$  it is monotonically increasing in the interval 
of $w$ satisfying $w_1 < w < w_{2}$ and monotonically decreasing 
in the interval of $w$ satisfying $w_2 < w< w_{3}$.
Therefore, if we derive $Z(w_{1})$, $Z(w_{2})$, and $Z(w_{3})$, 
we obtain the numbers of the two-string solutions for narrow and wide 
pairs, 
respectively. 

\subsection{Narrow pairs ($w_1 < w < w_2$)}

\subsubsection{Counting function $Z(w)$ at $w=w_2$ 
in the limit of $w\rightarrow w_{2}$ with $w < w_2$}\label{sec:unstablew2}

If $w$ is close to 1, we define $\epsilon$ as $\epsilon \equiv 1-w$. 
We expand $\widehat{A}$, $\widehat{B}$ and $\widehat{C}$ in terms of $\epsilon$.  
\begin{eqnarray}
\widehat{A}(1-\epsilon)
&= & 4 \left(\frac {1-t^2} 2 \right)^{2-2/N} - (1+t^2)^{2-2/N} \epsilon^{2/N} + \cdots 
\nonumber\\
\widehat{ B}(1-\epsilon)& = & - {\frac 4 {t^2}} \left( \frac {1-t^2} 2 \right)^{2-2/N} + 
\left( t^{-2} + 2 + 5 t^2 \right) \frac {\epsilon^{2/N}} {(1+t^2)^{2/N}} + \cdots  
\nonumber \\
\widehat{ C}(1-\epsilon)& = & 4 \frac {\epsilon^{2/N}} {(1+t^2)^{2/N}} + \cdots \, . 
\end{eqnarray}
In terms of $\widehat{A}(w)$ 's the function $X= \tan^2 x$ is expressed as  
\begin{equation}
X(w) \equiv \frac{1}{2 \widehat{A}}\Bigl(\widehat{ B} + 
\sqrt{\widehat{B}^2+ 4\widehat{A} \widehat{ C}}\Bigr) \, . \label{eq:X(w)}\\
\end{equation}
We thus have 
\begin{equation}
X(w) = \frac {t^2} {(1+t^2)^{2/N}} \left( \frac 2 {1-t^2} \right)^{2-2/N} \epsilon^{2/N}  + \cdots \, .  
\end{equation} 
It follows that  $X$ approaches 0 as $\epsilon$ goes to zero:  
$X(1-\epsilon) \rightarrow 0$ ( $\epsilon \rightarrow 0$). 

For $\tan x> 0$, quantities $a$ and $b$ are given in terms of $X$ by 
\begin{eqnarray} 
a &=&\frac{\sqrt{X} (1-w^2 t^2 )}{ t \left(1+ X^2 w^2 t^2 \right)} 
\label{eq:aX} \\
b &=&\frac{(1+X) w}{1+ X w^2 t^2 } \, .  
\label{eq:bX}
\end{eqnarray}
Furthermore, we can show  that $a \rightarrow 0 $ and $a > 0 $ for $w \rightarrow w_2$
\begin{eqnarray}
b = 1 + \frac {t^2 (1-t^2)} {(1+t^2)^{2/N}}  \left( \frac 2 {1-t^2}  \right)^{2-2/N} \epsilon^{2/N} 
+ \cdots \, .  \label{eq:bw2m}
\end{eqnarray} 
We therefore have that $b > 1$ for $w=1- \epsilon $ with $0< \epsilon \ll1$.  

Let us denote by $\lim_{w \downarrow c}$ the right-hand limit where we send $w$ with $w>c$. Likewise we denote by $\lim_{w \uparrow c}$ the left-hand limit where we send $w$ to $c$ 
with $w<c$.

In the case of $\tan x> 0$ for $w< w_2$ 
we have from (\ref{eq:bw2m})
\begin{eqnarray} 
& \frac a {1-b} \rightarrow  - \infty  \quad ( w \uparrow w_2 )  
\nonumber \\ 
& \frac a {1+b} \rightarrow 0  \quad ( w \uparrow w_2 )
\end{eqnarray}
We thus evaluate the counting function $Z(w)$ in eq. (\ref{eq:counting1})
\begin{eqnarray}
Z(w) 
&\rightarrow - {\frac 1 4} + 0+ {\frac 1 2} 
\biggl(1  \biggr)   \quad ( w \uparrow w_2 )
\nonumber \\ 
& \quad = {\frac 1 4} .
\end{eqnarray}
We thus have in the limit of $\sqrt{X}$ approaching zero while  
$w$ approaching 1, 
the counting function becomes $1/4$. It may lead to a singular solution.

\subsubsection{In the stable regime the counting function $Z(w)$ at $w=w_{1}$ 
$(0<w_{1})$ leading to extra two-string solutions}

We recall that $\widehat{C}(w_1) = 0$, as shown in Lemma 1.   
Furthermore, 
we can show that $\widehat{B}(w_1) > 0$.   Then, 
in the limit of sending $w$ to $w_1$ for $w$ with $w_1 < w$,  
we have  from (\ref{eq:X(w)})
\begin{eqnarray}
X(w) =
\rightarrow\infty \quad ( w \downarrow w_1 )  .
\end{eqnarray}
We thus have 
\begin{eqnarray} 
a &=&\frac{\sqrt{X} (1-w^2 t^2 )}{ t \left(1+ X^2 w^2 t^2 \right)}   
\rightarrow 0 \quad ( w \downarrow w_1 ) \, ,  \nonumber \\
b &=&\frac{(1+X) w}{1+ X w^2 t^2 }   
\rightarrow\frac{1}{ w_1 t^2 } > 1 \quad ( w \downarrow w_1 )  \, . 
\end{eqnarray}
In the case of $\tan x >0$ the 
counting 
function (\ref{eq:counting1}) approaches  
\begin{eqnarray}
Z(w) =&\rightarrow 0+0+ \frac 1 2 \, .  \quad ( w \downarrow w_1 )
\end{eqnarray}
We thus obtain $N Z(w_{1})$.
\begin{eqnarray}
N Z(w_{1})=\frac{N}{2}.
\end{eqnarray}

We suggest that there should be no corresponding solution to 
the limit $N Z(w_{1}) = N/2$,  since $X(w)$ goes to infinity at $w=w_1$.  
Due to the parity 
condition 
(\ref{eq:QN}), the largest 
quantum number is given by $(N-1)/2$. 

If there is a complex solution such that its quantum number is larger than the largest number predicted by the string hypothesis, we call it an extra two-string solution. 
According to the string hypothesis, the maximum Bethe quantum number $J_1$ should be given by $(N-3)/2$ for narrow pairs. (See in Appendix D.) 
However, even in the stable regime 
we have a complex solution for the Bethe quantum number  $J_1=(N-1)/2$, 
since we have $N Z(w_1) = N/2$.  
In conclusion,  the number of two-string solutions 
is larger than the number predicted by the string hypothesis throughout 
the whole stable regime.

\subsubsection{In the unstable regime 
the counting function at $w=0$  and collapse conditions}

In the unstable regime we have $w_{1}=0$.  
For $0 < w \ll1$ we can derive 
\begin{eqnarray}
X(w) & =  \frac {\widehat{C}(w)} {| \widehat{B}(w)|} + \cdots 
\nonumber \\ 
& = \frac{N - t^2 N_{\zeta}} {N_{\zeta} - N} + \cdots  \,.    
\end{eqnarray} 
Since $X$ approaches a constant it is easy to show  
\begin{eqnarray}
b \rightarrow 0 \quad (w \rightarrow 0 )
\end{eqnarray} 
We therefore calculate the counting function at $w=0$ as follows. 
\begin{eqnarray}
Z(w) =
&\rightarrow\frac{1}{2\pi}\biggl\{2\tan^{-1}\Biggl(\frac{1}{t} \sqrt{\frac{\widehat{ C}(0)}
{|\widehat{ B}(0)|} } \Biggr) 
\biggr\} \qquad  (w \rightarrow 0)\nonumber\\
&=\frac{1}{\pi}\tan^{-1}\Biggl(\sqrt{\frac{N-(1+t^{2})}{1-(N-1) t^{2})}}\Biggr) \, . 
\end{eqnarray}
We denote  $Z(0)$ also by $Z_0^{(\zeta, N)}$. 
In the limit of sending  $\zeta$ to 0 (i.e. the anisotropy parameter $\Delta \rightarrow 1$, 
the XXX limit) we have 
\begin{eqnarray}
\lim_{\zeta\rightarrow0} N Z_{0}^{(\zeta,N)} =\frac{N}{\pi}\tan^{-1}\bigl(\sqrt{N-1}\bigr).
\end{eqnarray}
It coincides with that of the XXX chain\cite{DG1}. 
The conditions that the collapse of $m$ two-string solutions occurs 
in the chain of $N$ sites for $m=1, 2, \ldots$, are given by  
\begin{eqnarray}
N Z_{0}^{(\zeta,N)} < \frac{N-(1+2m)}{2}\label{eq:Z0m}.
\end{eqnarray}
If there is a complex solution such that  its quantum number is equal to 
$(N-1)/2$, we call it  an extra two-string solution. 
The condition for an extra two-string solution appears  is given by 
\begin{eqnarray}
\frac{N-1}{2}< N Z_{0}^{(\zeta, N)}   \label{eq:Z0e}.
\end{eqnarray}
Therefore, we obtain the following results.\\
\bf{Conjecture 1}\rm{}
If $N$ and $\zeta$ satisfy
\begin{eqnarray}
\tanh^2 \left( \zeta/2 \right) <   
\frac{ 1- (N-1) \tan^{2} \Bigl(\pi(1+2m)/2N \Bigr)}
{(N-1) - \tan^{2}\Bigl(\pi(1+2m)/2N \Bigr) } \label{eq:m-collapse}
\end{eqnarray}
the collapse of $m$ two-string solutions occurs. \\
\bf{Conjecture 2}\rm{}
If $N$ and $\zeta$ satisfy
\begin{eqnarray}
\tanh^{2}\left( \zeta/2 \right) 
> \frac {1- (N-1) \tan^{2} \Bigl( \pi/2N \Bigr) }
{ (N-1) - \tan^{2} \Bigl( \pi/2N \Bigr) }  \label{eq:Z0e1}
\end{eqnarray}
an extra pair of two-string solutions appears. 

We have depicted the results in figures \ref{fig:extra-sol} and \ref{fig:missing-sol}.

\subsubsection{Bethe quantum numbers for narrow pairs}  (i) In the 
stable 
regime the Bethe quantum numbers $J_1$ for narrow pairs (for $w$ with $w_1 < w < w_2$) are given by   
\begin{eqnarray} 
 {\frac N 4}  \le J_1 < \frac{N} 2   
 \quad  (\tan x > 0)  \, , \nonumber \\ 
 -  {\frac{N} 2} < J_1 \le - {\frac N 4} \quad (\tan x < 0) . 
\label{eq:QNnarrow-stable}
\end{eqnarray} 
(ii) In the unstable regime  the Bethe quantum numbers $J_1$ for narrow pairs 
(for $w$ with $w_1 < w < w_2$) are given by   
\begin{eqnarray} 
 {\frac N 4}  \le J_1 \le \frac{N}{\pi}\tan^{-1}\Biggl(\sqrt{\frac{N-(1+t^{2})}{1-(N-1) t^{2})}}\Biggr) \ 
 \quad  (\tan x > 0)  \, , \nonumber \\ 
 -  \frac{N}{\pi}\tan^{-1}\Biggl(\sqrt{\frac{N-(1+t^{2})}{1-(N-1) t^{2})}}\Biggr) \le J_1 \le - {\frac N 4} \quad (\tan x < 0) . 
\label{eq:QNnarrow-unstable} 
\end{eqnarray}

\subsection{Wide pairs ($w_2 < w < w_3$) } 

\subsubsection{Counting function at $w=w_{3}$ }

We recall that $\widehat{C}(w_3) > 0$, as shown in Lemma 1.   
In fact, we can show it directly. 
Furthermore, 
we can show that $\widehat{B}(w_3) > 0$.  
Then, in the limit of sending $w$ to $w_3$ for $w$ with $w < w_3$,  
$\widehat{A}(w)$ approaches $0$ with $\widehat{A}(w)>0$, 
and hence we have  from (\ref{eq:X(w)})
\begin{eqnarray}
X(w) =
\rightarrow\infty \quad ( w \uparrow w_3 )  .
\end{eqnarray}
We thus have 
\begin{eqnarray} 
a &=& \rightarrow 0 \quad ( w \uparrow w_3 ) \, ,  \nonumber \\
b &=& \rightarrow\frac{1}{ w_3 t^2 } > 1 \quad ( w \uparrow w_3 )  \, . 
\end{eqnarray}
Here we have assumed that  $w_3 t^2 <1$. In fact, $w_3$ approaches 1 
exponentially as $N$ increases.  
We therefore have from (\ref{eq:counting1})
\begin{eqnarray}
Z(w) & 
&\rightarrow 0+0+\pi\biggl(1-\frac{1}{N}\biggr) \quad (w \rightarrow 1). 
\end{eqnarray}
We obtain $N Z(w_{3})$ as follows. 
\begin{eqnarray}
N Z(w_{3})=\frac{N-1}{2}.
\end{eqnarray}

We suggest that there should be no corresponding solution to 
the limit of the counting function: $N Z(w_{3}) = (N-1)/2$, 
 since $X(w)$ goes to infinity at $w=w_3$,   
although the parity 
condition 
(\ref{eq:QN}) holds for  $(N-1)/2$. 
The largest quantum number for wide pairs 
is given by  $J_1= (N-3)/2$.

\subsubsection{Counting function in the limit $w \rightarrow w_{2}$ for 
$w$ with $w_2 < w < w_3$ }\label{sec:w2}

When $w$ is close to 1,  we define $\epsilon$ as $\epsilon\equiv w-1$.
We expand $\widehat{A}$, $\widehat{B}$ and $\widehat{C}$ with $\epsilon$.  

\begin{eqnarray}
\widehat{A}(1 + \epsilon)
&= & 4 \left(\frac {1-t^2} 2 \right)^{2-2/N} - (1+t^2)^{2-2/N} \epsilon^{N/2} + \cdots 
\nonumber\\
\widehat{ B}(1 + \epsilon)& = & - {\frac 4 {t^2}} \left( \frac {1-t^2} 2 \right)^{2-2/N} + 
\left( t^{-2} + 2 + 5 t^2 \right) \frac {\epsilon^{2/N}} {(1+t^2)^{2/N}} + \cdots  
\nonumber \\
\widehat{ C}(1 + \epsilon)& = & 4 \frac {\epsilon^{2/N}} {(1+t^2)^{2/N}} + \cdots \, . 
\end{eqnarray}
We recall that function $X(w) (= \tan^2 x )$ is expressed in terms of  
$\widehat{A}(w)$ 's  as shown in eq. (\ref{eq:X(w)}). 
We have 
\begin{equation}
X(w) = \frac {t^2} {(1+t^2)^{2/N}} \left( \frac 2 {1-t^2} \right)^{2-2/N} \epsilon^{2/N}  
+ \cdots .  \label{eq:Xw2up}
\end{equation} 
We thus have $X(1 + \epsilon) \rightarrow 0$ as $\epsilon \rightarrow 0$. 
Furthermore, we can show  
\begin{eqnarray}
b = 1 + \frac {t^2 (1-t^2)} {(1+t^2)^{2/N}}  \left( \frac 2 {1-t^2}  \right)^{2-2/N} \epsilon^{2/N} 
+ \cdots \, .  
\end{eqnarray} 
We therefore have that $b > 1$ for $w=1+ \epsilon $ with $0< \epsilon \ll1$.

By making use of the expansion of $X(1+\epsilon)$ shown in eq. (\ref{eq:Xw2up})  we have
\begin{eqnarray}
\frac{a}{1-b}\rightarrow  -\infty \qquad  (\epsilon\downarrow 0)\\
\frac{a}{1+b}\rightarrow0 \qquad (\epsilon\downarrow 0).
\end{eqnarray}
We thus obtain $N Z(w_2)$ for wide pairs  ($w_2 < w < w_3$) 
\begin{eqnarray}
NZ (w)  & \rightarrow - {\frac {N} 4}+0+ {\frac N 2} \biggl(1-\frac{1}{N}\biggr) 
\quad (w \downarrow w_2) \nonumber \\ 
  & = {\frac N 4} - {\frac 1 2}
\end{eqnarray}

\subsubsection{Bethe quantum numbers for wide pairs}   

The Bethe quantum numbers $J_1$ for narrow pairs 
(for $w$ with $w_1 < w < w_2$) 
are given by   
\begin{eqnarray} 
 {\frac N 4} - {\frac 1 2} \le J_1 < {\frac {N-1} 2}  \quad  (\tan x > 0)  \, , 
\nonumber \\ 
 -{\frac {N+1} 2} < J_1 \le - {\frac N 4} -{\frac 1 2} \quad (\tan x < 0) . 
\label{eq:QNwide}
\end{eqnarray} 

It is interesting to note that  
the right-hand limit and the left-hand limit for the counting function $Z(w)$ at $w=w_2=1$are different. 
\begin{description}
\item[(1)] Sending $\delta$ to $0$ with $\delta>0$ : $\delta\downarrow 0$ 
\begin{eqnarray}
N Z(w_{2}) = 
&=\frac{N}{4}-\frac{1}{2}   \quad (w \downarrow w_2) 
  \label{eq:delta0(1)}
\end{eqnarray}
It corresponds to a singular solution in the case of $N=4n$ with an integer $n$. \\
\item[(2)] Sending $\delta$ to $0$ with $\delta<0$ : $\delta \uparrow 0$ 
\begin{eqnarray}
N Z(w)  
&=\frac{N}{4}  \quad (w \uparrow w_2) 
\label{eq:delta0(2)} 
\end{eqnarray}
It corresponds to a singular solution in the case of $N=4n+2$ with an integer $n$. 
\end{description}

\subsection{Quantum numbers of a singular solution }\label{sec:SS}

In the spin-1/2 XXX spin chain the singular solutions corresponds to the case of 
$\delta=0$ and $x=0$ for two down-spins. 
Here we recall that $\delta=0$ corresponds to $w=1$. 

We now show that when $w=1$, we have $x=0$.
In fact, it has already been shown in section 2.4. 


It follows that when $J_{1}=Z(w=1)$, $J_1$ corresponds to a Bethe quantum number 
of the singular solution with $x=0$ and $\delta=0$.
From (\ref{eq:delta0(1)}) and (\ref{eq:delta0(2)}), when $N=4n$ with an integer n, the Bethe quantum numbers are $(J_{1}, J_{2})=(N/4-1/2,N/4+1/2)$ and $(-N/4-1/2,-N/4+1/2)$. Similarly, when $N=4n+2$ with an integer n, the Bethe quantum numbers are $(J_{1}, J_{2})=(N/4,N/4)$ and $(-N/4,-N/4)$.

%
%
%
%

\section{Proof of complex solutions approaching complete strings exponentially 
with respect to $N$
}\label{sec:RSH}

\subsection{Definition of functions $f$ and $g$ and two lemmas}

Let us define $f^{(N,\zeta)}(w)$ and $g^{(N,\zeta)}(w)$ by 
\begin{eqnarray}
f^{(N,\zeta)}(w)\equiv \Biggl\{\biggl(\frac{1+w t^{2}}{1-w t^{2}}\biggr)^{2}
\Biggr\}^{1-\frac{1}{N}}\\
g^{(N,\zeta)}(w)\equiv \Biggl\{\biggl(\frac{1+w}{1-w}\biggr)^{2}\Biggr\}^{\frac{1}{N}}.
\end{eqnarray}
It is easy to 
show 
that the equation for $w$: $\widetilde{ A}(w)=0$  
is equivalent to the equation for $w$: $f^{(N,\zeta)}(w)= g^{(N,\zeta)}(w)$.  
Furthermore,  
$\widetilde{ A}(w) < 0$ is equivalent to  $f^{(N,\zeta)}(w) >  g^{(N,\zeta)}(w)$, 
 for $w$ in  $0 < w < 1/t$.

\par \noindent 
{\bf{Lemma 4}} \\
\rm{} 
For any given pair of integers satisfying $N_{2}>N_{1}(>1)$, we have 
\begin{eqnarray}
g^{(N_{1},\zeta)}(w)>g^{(N_{2},\zeta)}(w)\ \ 
\mbox{for}\ 0<w<1\ \mbox{and}\ 1<w<1/t. 
\end{eqnarray}
{\bf{Proof}}\\
It simply follows from the inequality:  $\left((1+w)/(1-w) \right)^{2}>1$.

\par \noindent 
{\bf{Lemma 5}} \\
\rm{} 
For given two 
integers 
satisfying $N_{2}>N_{1}(>1)$, we have 
\begin{eqnarray}
f^{(N_{2},\zeta)}(w)>f^{(N_{1},\zeta)}(w)\ \ \mbox{for}\ 0<w<1/t . 
\end{eqnarray}
{\bf{Proof}}\\
It follows from  $(1+wt^{2})/(1-wt^{2})>1$. 

\subsection{Large-N behavior of the deviation in the interval $0<w<1$}\label{sec:0<1}

\begin{figure}
\begin{center}
\includegraphics[width=12cm]{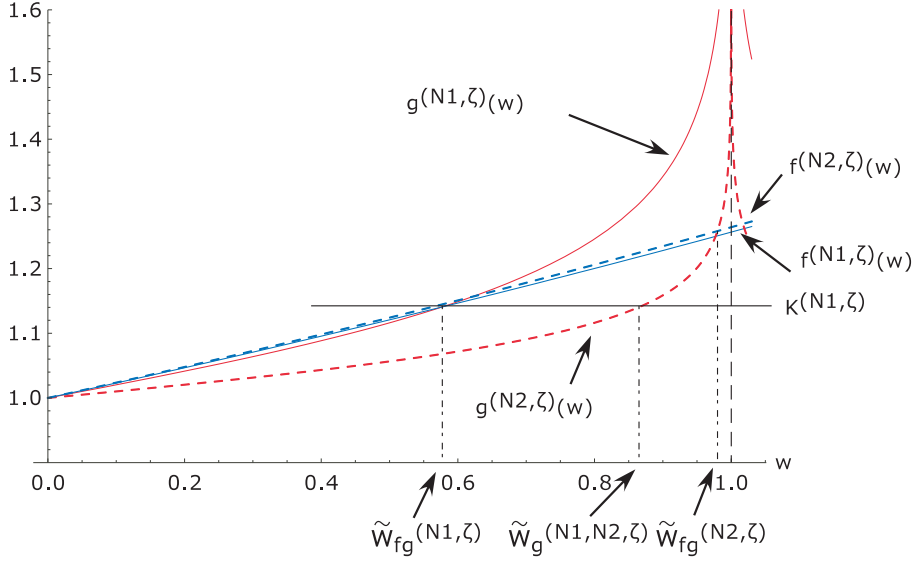}
\caption{In the case of narrow pairs: for $w$ satisfying $0<w<1$ and 
in the stable regime: $N > N_{\zeta}$. 
Crossing point  $\tilde{w}_{fg}^{(N_2, \zeta)}$ 
of $f^{(N_2, \zeta})$ and $g^{(N_2, \zeta)}$ 
is much closer to 1 than 
crossing point  $\tilde{w}_{fg}^{(N_1, \zeta)}$ 
of $f^{(N_1, \zeta)}$ and $g^{(N_1, \zeta)}$ for $N_1 < N_2$. 
}
\end{center}
\end{figure}

For narrow pairs, we assume that $N$ and $\Delta$ (or $\zeta$) 
are in the stable regime.  
Let us take a pair of integers $N_{1}$ and $N_{2}$ with $N_{2}>N_{1}>1$.  
First for $N_1$, 
we consider two graphs: $y=f^{(N_{1},\zeta)}(w)$  and $y= g^{(N_{1},\zeta)}(w)$. 
It has been shown in section \ref{sec:A0} that 
$\widetilde{ A}(w)< 0$ near the origin $w=0$ for the stable regime, 
and also that the two graphs  have the unique crossing point at $w=w_1$,    
which has been defined as the zero of $\widetilde{ A}(w)$: $\widetilde{ A}(w_1)=0$. 
We denote $w_1$ by $\tilde{w}_{fg}^{(N_{1},\zeta)}$, 
since  it is also a solution of the equation: $f^{(N_{1},\zeta)}(w)=g^{(N_{1},\zeta)}(w)$. 
We then define $K^{(N_{1},\zeta)}$ by the value of the function $f^{(N_1, \zeta)}$ at $w=w_1= \tilde{w}_{fg}^{(N_{1},\zeta)}$  
\begin{eqnarray}
K^{(N_{1},\zeta)}\equiv f^{(N_{1},\zeta)}(\tilde{w}_{fg}^{(N_{1},\zeta)}) 
\, \left( =g^{(N_{1},\zeta)}(\tilde{w}_{fg}^{(N_{1},\zeta)}) \right).
\end{eqnarray}
Here we remark that $y= K^{(N_{1},\zeta)}$ is a linear graph, 
and the graph of $y=g^{(N_2,\zeta)}(w)$ is rapidly monotone increasing 
with respect to $w$, which goes to infinity at $w=1$.  
We denote by $\tilde{w}_{g}^{(N_{1},N_{2},\zeta)}$ 
the crossing point of the two graphs $y=K^{(N_{1},\zeta)}$ 
and $y=g^{(N_{2},\zeta)}(w)$. 

Through the expansion of the logarithmic functions 
$\log(1 \pm z)$ it is easy to show  
\begin{equation}
\frac {1+ z} {1-z} > \exp( 2 z) \, \quad (-1 <  z < 1) \,. 
\label{eq:ineq} 
\end{equation}
We thus estimate the ratio with an exponential  
\begin{eqnarray}
& K^{(N_1, \zeta)}  = f^{(N_1, \zeta)}(\tilde{w}_{fg}^{(N_1, \zeta)}) 
\nonumber \\ 
&   = \left( \frac {1 + \tilde{w}_{fg}^{(N_1, \zeta)} t^2}
                        {1 - \tilde{w}_{fg}^{(N_1, \zeta)} t^2}   \right)^{2- 2/N_1}  
> \exp\left(    4 \tilde{w}_{fg}^{(N_1, \zeta)} t^2 \left(1 - 1/N_1 \right)  \right)  
\end{eqnarray}
We therefore express $\tilde{w}_{g}^{(N_{1},N_{2},\zeta)}$ 
with respect to $N_{2}$ when it is very large  
\begin{eqnarray}
\ \ \ \ \ g^{(N_{2},\zeta)}(\tilde{w}_{g}^{(N_{1},N_{2},\zeta)})=
K^{(N_{1},\zeta)}\nonumber\\
\Leftrightarrow\biggl(\frac{1+\tilde{w}_{g}^{(N_{1},N_{2},\zeta)}}{1-\tilde{w}_{g}^{(N_{1},N_{2},\zeta)}}\biggr)=
(K^{(N_{1},\zeta)})^{\frac{N_{2}}{2}}\nonumber\\
\Leftrightarrow \tilde{w}_{g}^{(N_{1},N_{2},\zeta)}
=\frac{1 - (K^{(N_{1},\zeta)})^{-\left( N_{2}/2 \right) }} 
{1+(K^{(N_{1},\zeta)})^{-\left( N_{2}/{2} \right) }} \nonumber\\
\qquad \qquad < 1 -2 \exp \left( - \alpha(N_1, \zeta) N_{2} \right) \,  ,  
\end{eqnarray}
where $\alpha(N_1, \zeta)$ is given by 
\begin{equation}
\alpha(N_1, \zeta) = 2 \tilde{w}_{fg}^{(N_1, \zeta)} \left(1 - 1/N_1 \right)  t^2 \,.  
\end{equation} 
It follows that $\tilde{w}_{g}^{(N_{1},N_{2},\zeta)}$ approaches 1 
exponentially with respect to $N_2$ as $N_{2}$ increases infinitely. 
Here we recall that $0< \tilde{w}_{fg}^{(N_1, \zeta)} $ 
since it corresponds to the zero $w_1$ of $\widetilde{A}(w)$ in the interval 
$0 < w < 1$.   

Furthermore, 
since $g^{(N,\zeta)}(w)$ and $f^{(N,\zeta)}(w)$ are monotone increasing with respect to $w$ and Lemma 5 holds, 
we have  $K^{(N_{1},\zeta)}<f^{(N_{1},\zeta)}(w)<f^{(N_{2},\zeta)}(w)$ 
for $w$ satisfying  $\tilde{w}_{fg}^{(N_{1},\zeta)}<w<1$. 
Because of this inequality, Lemma 4
and the fact that $g^{(N,\zeta)}(w)$ is monotone increasing with respect to $w$, 
we have $\tilde{w}_{g}^{(N_{1},N_{2},\zeta)} <  \tilde{w}_{fg}^{(N_{2},\zeta)}$. 
Since $\tilde{w}_{g}^{(N_{1},N_{2},\zeta)}$ approaches 1 exponentially 
fast with respect to $N_2$ as $N_2$ increases, it follows that 
$\tilde{w}_{fg}^{(N_{2},\zeta)}$ also becomes close to 1 
exponentially fast with respect to $N_2$.

Here we give a remark.  
If we employ $N_2$ as $N_1^{'}$ and take another number larger than $N_1^{'}$ 
as $N_2^{'}$,  the value $\alpha(N_1^{'}, \zeta)$ becomes closer to $2 t^2$  
than $\alpha(N_1, \zeta)$. 
We therefore have a conjecture  that the value $\alpha(N_1, \zeta)$ approaches 
$2 t^2$ as $N_1$ increases. 
That is, we expect that  
$\tilde{w}_{fg}^{(N_{1},\zeta)} (1-1/N_1)$ approaches 1 as 
$N_1$ goes to infinity. Consequently, we also expect that  
$\tilde{w}_{fg}^{(N,\zeta)} < 1 - 2 \exp \left( - 2 t^2 N \right)$ if 
 $N$ is large enough.

\subsection{Large-N behavior of the deviation in $1<w<1/t$}\label{sec:1<w}

For wide pairs we consider both the stable and unstable regimes.  

Let us take a pair of integers $N_{1}$ and $N_{2}$ satisfying $N_{2}>N_{1}>1$. 
In order to express the values of the function $f^{(N,\zeta)}(w)$ at $w=1$ for 
$N=N_1$ and $N=N_2$,   we define $K_1^{(N, \zeta)}$ by  
\begin{eqnarray}
K_{1}^{(N,\zeta)}\equiv f^{(N,\zeta)}(w=1). \label{eq:def-K1}
\end{eqnarray}
First, we consider two graphs: $y=K_{1}^{(N_{1},\zeta)}$ and $y=g^{(N_{2},\zeta)}(w)$.  
The former is a  horizontally flat line and the latter is 
monotonically 
decreasing with respect to $w$ and it comes from 
infinity at $w=1$.  
It is therefore clear that the two graphs have a crossing point at some $w$.  
We denote it by $w_{g}^{(N_{1},N_{2},\zeta)}$. 
Secondly, we consider another pair of graphs:  
$y=f^{(N_{2},\zeta)}(w)$ and $y=g^{(N_{2},\zeta)}(w)$.
For wide pairs ($1 < w < 1/t$) 
the function $y=f^{(N,\zeta)}(w)$ is monotone increasing with respect to $w$, 
while $y=g^{(N,\zeta)}(w)$ is monotone decreasing with respect to $w$.
It is therefore clear that they have a crossing point at some $w$, and we denote it 
by $w_{fg}^{(N_{2},\zeta)}$.

It follows from Lemma 5 that we have
\begin{eqnarray}
K_{1}^{(N_{1},\zeta)}=f^{(N_{1},\zeta)}(w=1)<K_{1}^{(N_{2},\zeta)}=f^{(N_{2},\zeta)}(w=1)
\end{eqnarray} 
Since $f^{(N_{2},\zeta)}(w)$ is monotone increasing we have 
\begin{eqnarray}
f^{(N_{2},\zeta)}(w=1) <f^{(N_{2},\zeta)}(w_{fg}^{(N_{2},\zeta)})=g^{(N_{2},\zeta)}(w_{fg}^{(N_{2},\zeta)})\label{eq;k1fg2}
\end{eqnarray}
From (\ref{eq;k1fg2}) and $g^{(N_{2},\zeta)}(w)$ is monotone decreasing, 
\begin{eqnarray}
1<w_{fg}^{(N_{2},\zeta)}<w_{g}^{(N_{1},N_{2},\zeta)}\label{relation1}
\end{eqnarray}

\begin{figure}
\begin{center}
\includegraphics[width=12cm]{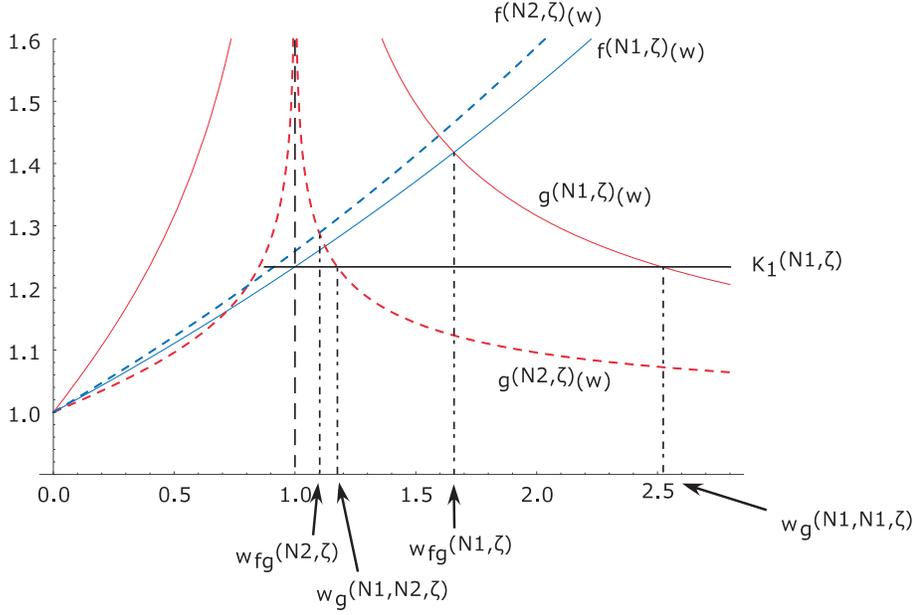}
\caption{In the case of wide pairs: for $w$ satisfying $1<w<1/t$ and 
both in the stable and unstable regimes. 
Crossing point  ${w}_{fg}^{(N_2, \zeta)}$ 
of $f^{(N_2, \zeta})$ and $g^{(N_2, \zeta)}$ 
is much closer to 1 than 
crossing point  ${w}_{fg}^{(N_1, \zeta)}$ 
of $f^{(N_1, \zeta)}$ and $g^{(N_1, \zeta)}$ for $N_1 < N_2$. 
}
\end{center}
\end{figure}

We now consider $g^{(N_{2},\zeta)}(w_{g}^{(N_{1},N_{2},\zeta)})$. 
We express $w_{g}^{(N_{1},N_{2},\zeta)}$ with $N_{2}$ as follows.
\begin{eqnarray}
&g^{(N_{2},\zeta)}(w_{g}^{(N_{1},N_{2},\zeta)})=K_{1}^{(N_{1},\zeta)}
\nonumber\\
&\Leftrightarrow \Bigl(\frac{w_{g}^{(N_{1},N_{2},\zeta)}+1}
{w_{g}^{(N_{1},N_{2},\zeta)}-1}\Bigr)^{2}= \left( K_{1}^{(N_{1},\zeta)} \right)^{N_{2}}
\nonumber\\
&\Leftrightarrow w_{g}^{(N_{1},N_{2},\zeta)}
=\frac{1+ \left( K_{1}^{(N_{1},\zeta)} \right)^{-N_{2}/2}} 
{1-  \left( K_{1}^{(N_{1},\zeta)} \right)^{-N_{2}/2} }\nonumber\\
& \qquad \qquad < 1+2 \left( K_{1}^{(N_{1},\zeta)} \right)^{-N_{2}/2}
\end{eqnarray}
Here by applying inequality (\ref{eq:ineq}) to $K_1^{(N, \zeta)}$ 
with $w=1$ we have 
\begin{eqnarray}
K_1^{(N, \zeta)}  & = f^{(N, \zeta)}(w=1) 
\nonumber \\ 
&   = \left( \frac {1 +  t^2}
                        {1 -  t^2}   \right)^{2- 2/N_1}  
> \exp\left(    4 t^2 \left(1 - 1/N \right)  \right)  
\end{eqnarray}
We therefore have 
\begin{equation}
w_{g}^{(N_{1},N_{2},\zeta)} < 1 + 2 \exp\left(- 2 t^2 \left(1- 1/N_1 \right) N_2  
 \right) \, .  
\end{equation}

Since $w_{g}^{(N_{1},N_{2},\zeta)}$ exponentially converges to 1 as $N_2$ increases 
infinitely 
(i.e., $N_{2}\rightarrow\infty$) 
and inequality (\ref{relation1}) holds,  it follows that 
$w_{fg}^{(N_{2},\zeta)}$ exponentially converges to 1 as 
$N_2$ increases 
infinitely 
(i.e. $N_{2}\rightarrow\infty$).

%
%
%
%

\section{Numerical solutions of two-strings } 
\label{sec:numerical}
\subsection{Behavior of two-strings as the XXZ anisotropy $\Delta$ approaches 1}

Two-string solutions do not collapse to real solutions if the XXZ anisotropy parameter $\Delta$ is large enough. The stable regime is located in the upper part of the whole diagram, i.e., in the area for large $N$ and large $\Delta$, 
as shown in figures \ref{fig:stable-unstable}, \ref{fig:extra-sol} and \ref{fig:missing-sol}.    

\begin{figure}[htb]
\includegraphics[width=0.9\columnwidth]{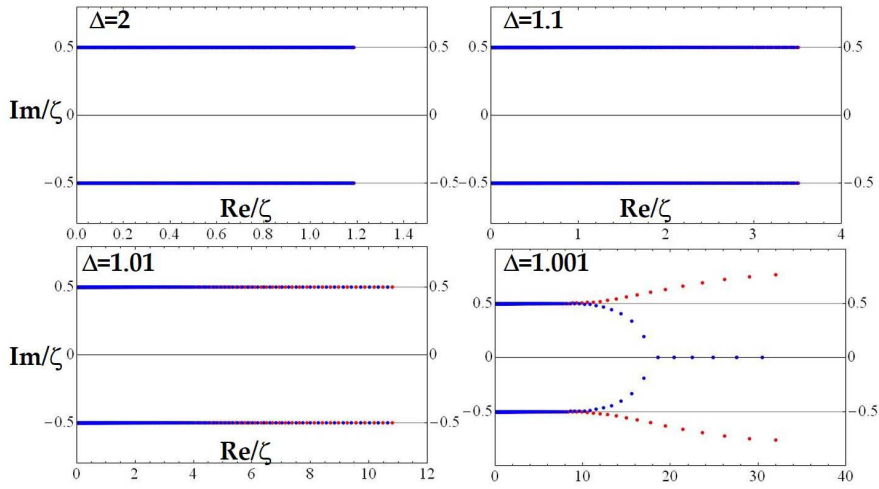}
\caption{ 
Rapidities of two-string solutions in the spin-1/2 massive XXZ chain 
with $N=1000$ in the two down-spin sector ($M=2$) 
for anisotropy parameter $\Delta=2, 1,1, 1.01$ and 1.001. 
Narrow pairs are depicted by blue dots, while wide pairs by red dots.}
\label{fig:delta}
\end{figure}

For an illustration, we plot the rapidities of two-string solutions 
with $N=1000$, 
i.e. the numerical solutions of the Bethe-ansatz equations (\ref{eq:BAE11}) 
and (\ref{eq:BAE22}) for $N=1000$,  
  in figures \ref{fig:delta} for  $\Delta=2$, 1.1,  1.01 and 1.001 .    
It is clear that string deviations become more significant as the anisotropy parameter $\Delta$ decreases to 1. 
In the case of $\Delta=1.001$ string deviations are large and nontrivial.

In the lower right panel of figure \ref{fig:delta}  we observe 
not only wide and narrow pairs with large deviations but also  
real solutions which correspond to collapsed two-string solutions.  

We observe six points on the real axis in the lower right panel 
of figure \ref{fig:delta}.  Let us analytically derive the number. 
From inequality (\ref{eq:m-collapse})  we derive an estimate of  
the number of collapsed two-string solutions for a given $N$ and $\zeta$ 
as follows.  
\begin{eqnarray}
n_{\rm collapse} < \frac N{\pi} \tan^{-1}\left( \sqrt{\frac {1- (N-1) t^2}  {N-1-t^2}}
\right) - \frac 1 2  \,. \label{eq:n_collapse}
\end{eqnarray} 

Let us consider the case of $N=1000$ and $\Delta =1.001$. 
For $\Delta =1.001$ we have $\zeta= \cosh^{-1}(1.001) =  0.0447176$ and 
then $t=\tanh(\zeta/2)= 0.0223551$. We thus have an estimate 
$n_{\rm collapse} < 6.62533$, i.e., we have  $n_{\rm collapse}=6$. 
We therefore have shown that the collapse of six two-string solutions occurs 
for $N=1000$ and $\Delta =1.001$.

\subsection{Behavior of two-strings as $N$ becoming very large}

\begin{figure}
\includegraphics[width=0.9\columnwidth]{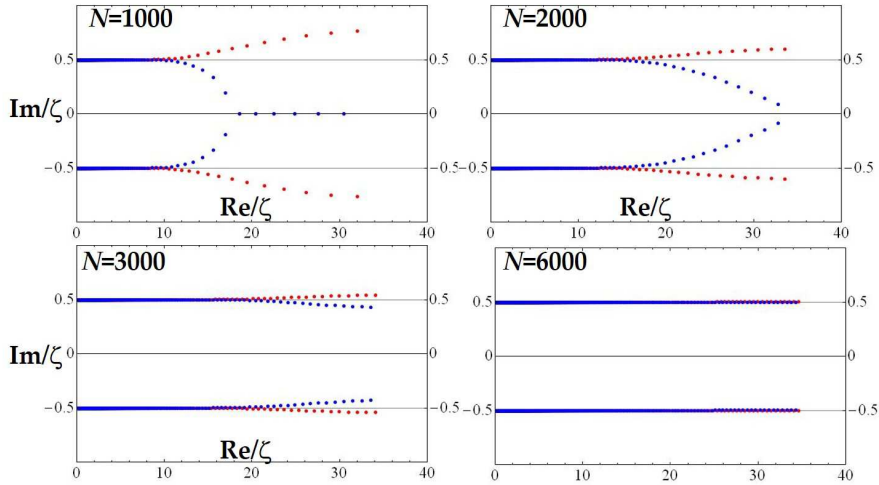}
\caption{ 
Rapidities of two-string solutions in the two down-spin sector ($M=2$) 
of the spin-1/2 massive XXZ chain 
with anisotropy parameter $\Delta=1.001$  
for  $N=1000$, 2000, 3000, and 6000. 
}
\label{fig:Nsite}
\end{figure}

We plot the rapidities of two-string solutions  
in figure \ref{fig:Nsite} for $N=$1000, 2000, 3000, and 6000. 
We observe that the string deviations become smaller as the number of sites $N$ increases, although $\Delta$ is the same and 
very close to 1 such as $\Delta=1.001$. 
We observe the behavior that the string deviations $\delta$ become smaller 
as the number of site $N$ increases, and they finally vanish for 
$N=6000$.

The observed numerical behavior is consistent with the theoretical result. 
We recall that in section 5 we have shown that the string deviations decreases 
exponentially with respect to the number of sites $N$.     

In the case of $N=2000$ and $\Delta =1.001$,  
from formula (\ref{eq:n_collapse}) we have 
$n_{\rm collapse} = -0.05$. 
We therefore have no collapse of complex solution for this case. 
It is consistent with the numerical result that no
narrow pairs are on the real axis in the

\subsection{Numerical confirmation  of extra 2-string solutions}

For an illustration,  we present an example of numerical two-string solutions 
in the case of $N=12$.  
It follows from condition (\ref{eq:Z0e1}) that 
if the anisotropy parameter $\zeta$ satisfies $0.556888\cdots<\zeta$, 
an extra pair of complex solutions of the Bethe ansatz equations 
appears. 

In table \ref{tb:2string} we give all the complex solutions of the Bethe-ansatz 
equations in  the massive regime of the spin-1/2 XXZ Heisenberg spin chain 
with $N=12$ sites and $\zeta=0.6$ (i.e., $\Delta=\cosh(0.6)=1.18546521824226770375\cdots$). 
Here we remark that it is in the unstable regime. 
\begin{equation}
N_{\zeta}= (1- t^2)/t^2 \approx 10.7837 \, . 
\end{equation} 
The energy eigenvalues corresponding to the solutions are consistent with the  estimates of the numerical diagonalization of the XXZ Hamiltonian. The solutions of No. 1 and No. 6 in table \ref{tb:2string}
are the extra pair of complex solutions. 
They are  not predicted by the string hypothesis.  
We remark that No.9 is the singular solution.

Let us consider the Bethe quantum numbers, explicitly. 
For wide pairs ($w_2 < w < w_3$), we have the upper and lower bounds 
(\ref{eq:QNwide}) for the quantum number $J_1$. 
\begin{eqnarray} 
& {\frac N 4} - {\frac 1 2} \le J_1 < \frac {N-1} 2 
\quad (\tan x > 0) \, , \nonumber \\
& - \frac {N+1} 2 <  J_1 \le - {\frac N 4} - {\frac 1 2} 
\quad (\tan x \le 0 ) \, .  \nonumber 
\end{eqnarray} 

In the case of $ x> 0$ for $N=12$ we have $(N-1)/2= 11/2$, 
and hence the largest quantum number  
$J_1$ is given by $J_1=9/2$. 
We recall that $J_2$ is given by $J_2=J_1 +1$ for wide pairs. 
For standard wide pairs with $x > 0$ we have  
\begin{equation}
(J_1, J_2) = (7/2, 9/2) \, , \quad (9/2, 11/2) \, . 
\end{equation}   
The smallest positive quantum number is 
given by $J_1= 5/2$, since $N/4 -1/2 = 5/2$. 
It corresponds to the singular solution, 
since 5/2 coincides with the lower bound. 

In the case of $ x< 0$ for $N=12$  the smallest quantum number $J_1$ 
is given by $J_1=-11/2$, since  $- (N+1)/2 = -13/2$.  
For standard wide pairs with $x < 0$ we have 
\begin{equation}
(J_1, J_2) = (-9/2, -7/2) \, , \quad (-11/2, -9/2) \, . 
\end{equation}   
The largest negative quantum number $J_1$ is 
given by $J_1= -7/2$, since we have $-N/4 -1/2 = -7/2$. 
It corresponds to the singular solution. 
Hence,  the two sets of quantum numbers correspond to the 
same singular solution $(x, \delta)=(0,0)$: 
\begin{equation}
(J_1, J_2) = (5/2, 7/2) \, , \quad (-7/2, -5/2) \, . 
\end{equation}

\begin{center}
\begin{table}[htb] 
  \begin{tabular}{c c c c}\hline
  No. & $J$ & $\lambda$& Energy  \\ \hline \hline

    \scriptsize1&  \scriptsize11/2 &   \scriptsize1.13537646891480325577+0.16312176718062221300i &   \scriptsize
$-$0.42692157141886207577 \\ 
     &  \scriptsize11/2 &     \scriptsize1.13537646891480325577$-$0.16312176718062221300i &  \\ \hline

    \scriptsize2& \scriptsize9/2 &  \scriptsize0.49443316603739513350+0.29840352572689955991i
 &   \scriptsize-0.76659148423211189359 \\ 
     & \scriptsize9/2  &\scriptsize0.49443316603739513350$-$0.29840352572689955991i
 &  \\ \hline

    \scriptsize3& \scriptsize7/2 &  \scriptsize0.14292089534049196825+0.29999999114716871863i  &  \scriptsize$-$1.12895830143233665108 \\ 
     & \scriptsize7/2 & \scriptsize0.14292089534049196825-0.29999999114716871863i  &  \\ \hline

    \scriptsize4& \scriptsize-7/2 &  \scriptsize$-$0.14292089534049196825+0.29999999114716871863i &  \scriptsize$-$1.12895830143233665108 \\ 
     & \scriptsize-7/2 & \scriptsize$-$0.14292089534049196825$-$0.29999999114716871863i  &  \\ \hline

\scriptsize5& \scriptsize-9/2 &  \scriptsize$-$0.49443316603739513350+0.29840352572689955991i
 &\scriptsize$-$0.76659148423211189359 \\ 
     & \scriptsize-9/2 &\scriptsize$-$0.49443316603739513350$-$0.29840352572689955991i
&  \\ \hline

\scriptsize6& \scriptsize-11/2 & \scriptsize$-$1.13537646891480325577+0.16312176718062221300i & \scriptsize-0.42692157141886207577\\ 
     & \scriptsize-11/2 &  \scriptsize$-$1.13537646891480325577$-$0.16312176718062221300i &  \\ \hline

    \scriptsize7&  \scriptsize11/2 &   \scriptsize0.74045039986314916894+0.31469282216447499146i  &   \scriptsize$-$0.54303402832662696979 \\ 
     &  \scriptsize9/2 &   \scriptsize0.74045039986314916894$-$0.31469282216447499146i  &  \\ \hline

    \scriptsize8& \scriptsize9/2 &  \scriptsize0.30062425150856577406+0.30002387970572190065i &  \scriptsize$-$0.97443500257666539810\\ 
     & \scriptsize7/2 &  \scriptsize0.30062425150856577406$-$0.30002387970572190065i &  \\ \hline

 \scriptsize9& \scriptsize-5/2 &  \scriptsize0.3i &  \scriptsize$-$1.18546521824226770375 \\ 
     & \scriptsize-7/2 &  \scriptsize$-$0.3i  &  \\ \hline

 \scriptsize10& \scriptsize$-$7/2 &  \scriptsize$-$0.30062425150856577406+0.30002387970572190065i  &  \scriptsize$-$0.97443500257666539810 \\ 
     & \scriptsize-9/2  & \scriptsize$-$0.30062425150856577406$-$0.30002387970572190065i &  \\ \hline

 \scriptsize11& \scriptsize-9/2  & \scriptsize$-$0.74045039986314916894+0.31469282216447499146i  &  \scriptsize$-$0.54303402832662696979 \\ 
     & \scriptsize-11/2 &  \scriptsize$-$0.74045039986314916894$-$0.31469282216447499146i   &  \\ \hline
  \end{tabular}
\caption{Numerical two-string solutions of the Bethe ansatz equations 
for $N=12$ in the two down-spin sector, at $\zeta=0.6$. 
No.1 and No.6 are extra complex solutions, which are not 
predicted by the string hypothesis. No.9 is the singular solution. It has 
another set of quantum numbers (5/2, 7/2). }
\label{tb:2string}
\end{table}
\end{center}

For narrow pairs ($w_1 < w < w_2$) in the unstable regime, 
we have the upper and lower bounds 
(\ref{eq:QNnarrow-unstable}) for the quantum number $J_1$. 
\begin{eqnarray} 
& {\frac N 4}  \le J_1 \le  Z_0^{(N=12,\zeta=0.6)} \approx 5.70244 
\quad  (\tan x > 0 ) \, , \nonumber \\
&  -Z_0^{(N=12, \zeta = 0.6)} \le  J_1 \le - {\frac N 4}  \quad  (\tan x < 0 ) \, . 
\end{eqnarray} 

In the case of $ x> 0$. 
for $N=12$ the largest quantum number  
$J_1$ is given by $J_1=11/2$, since we 
have 
$Z_0^{(N=12,\zeta=0.6)} \approx 5.70244 $.  
We recall that  $J_2=J_1$ for narrow pairs. 
The smallest positive quantum number is given by $J_1= 7/2$,  
since we have $N/4 = 3$.  
For narrow pairs with $x > 0$ we have 
\begin{equation}
(J_1, J_2) = (7/2, 7/2) \, , \quad (9/2, 9/2) \, , \quad (11/2, 11/2) \, . 
\end{equation}   
Here the set (11/2, 11/2) gives an extra two-string solution, while the first two sets 
of quantum numbers are standard in the string hypothesis.  

In the case of $ x< 0$ for $N=12$  the smallest quantum umber $J_1$ 
is given by $J_1=-11/2$, since we have $- N/2 = -6$. 
The largest negative quantum number $J_1$ is 
given by $J_1= -7/2$, since we have $-N/4  = -3$. 
 For narrow pairs with $x < 0$ we have 
\begin{equation}
(J_1, J_2) = (-11/2, -11/2) \, , \quad (-9/2, -9/2) \, , \quad (-7/2, -7/2) \, . 
\end{equation}   
Here the set (-11/2, -11/2) gives an extra two-string solution.

\section{The XXX/XXZ correspondence of the Bethe quantum number}\label{sec:XXXXXZ}

We now argue that if there exists a finite-valued solution of the Bethe ansatz equations for the spin-1/2 XXX spin chain, 
then  there exists a solution of the Bethe ansatz equations for the spin-1/2 XXZ spin chain such that it has the same set of the Bethe quantum numbers
as the XXX solution and  it converges to the XXX solution in the limit of 
sending the anisotropy parameter $\zeta$ to zero.  

We shall give the argument particularly in the sector of two down-spins. 
However,  we expect that the statement should be valid in any given sector. 
It is the merit of considering the two down-spin sector of the spin 1/2 XXX spin chain that it as been shown rigorously \cite{DG1} that all the solutions are associated with the different set of the Bethe quantum numbers and the solutions 
 are distinct.

The argued correspondence between the solutions of the XXX spin chain and those of the XXZ spin chain should be fundamental.  Here we remark that 
some of the finite-valued solutions of the XXZ spin chain become divergent 
in the limit of sending $zeta$ to zero.  It is due to the fact that  
the total spin SU(2) symmetry of the XXX spin chain is broken in the XXZ spin chain.

As a reference we present  
the logarithmic form of the Bethe ansatz equations in the 
spin-1/2 XXX spin chain ($\Delta=1$) for $M$ down-spins. 
\begin{eqnarray}
2\tan^{-1}(2\lambda_{i})=\frac{2\pi}{N}J_{i}+\frac{1}{N}\sum_{k=1}^{M}2\tan^{-1}(\lambda_{i}-\lambda_{k})\\
J_{i}\equiv\frac{1}{2}(N-M+1)\ \ \ (\mbox{mod} 1)  \quad (j=1, 2, \ldots, M).
\end{eqnarray} 
Hereafter we set $M=2$.  

Let us assume that we have a pair of rapidities $\lambda_1$ and $\lambda_2$ 
that satisfy the Bethe ansatz equations (\ref{eq:BAE11}) and (\ref{eq:BAE22}) 
with the Bethe quantum numbers $J_1$ and $J_2$, respectively.  
We then define reduced rapidities $\bar{\lambda_{1}}$ and  $\bar{\lambda_{2}}$, respectively,  by 
\begin{eqnarray}
\bar{\lambda_{1}}\equiv\frac{\lambda_{1}}{\zeta},\ \ \bar{\lambda_{2}}\equiv\frac{\lambda_{2}}{\zeta}  \, . 
\label{barlambda}
\end{eqnarray}
We substitute rapidities $\lambda_1$ and $\lambda_2$ of 
eqs. (\ref{barlambda}) into the Bethe ansatz equations (\ref{eq:BAE11}). We first calculate the left hand-side of eq. (\ref{eq:BAE11}) 
when $\zeta$ is very small, as follows. 
\begin{eqnarray}
\frac{\tan(\lambda_{1})}{\tanh(\zeta/2)}&=\frac{\tan(\zeta\bar{\lambda_{1}})}{\tanh(\zeta/2)}\nonumber\\
&\approx\frac{(\zeta\bar{\lambda_{1}})+\frac{1}{3}(\zeta\bar{\lambda_{1}})^{3}}{(\zeta/2)-\frac{1}{3}(\zeta/2)^{3}}\nonumber\\
&\approx2\bar{\lambda_{1}}+\frac{2}{3}(\bar{\lambda_{1}})^{3}\zeta^{2}\ \ (\zeta\ll1)\label{cal1}
\end{eqnarray}
We then calculate the main part in the right hand-side of eq. (\ref{eq:BAE11}) as 
\begin{eqnarray}
\frac{\tan(\lambda_{1}-\lambda_{2})}{\tanh(\zeta)}&=\frac{\tan\bigl(\zeta(\bar{\lambda_{1}}-\bar{\lambda_{2}})\bigr)}{\tanh(\zeta)}\nonumber\\
&\approx\frac{\zeta(\bar{\lambda_{1}}-\bar{\lambda_{2}})+\frac{1}{3}\zeta^{3}(\bar{\lambda_{1}}-\bar{\lambda_{2}})^{2}}{\zeta-\frac{1}{3}\zeta^{3}}\nonumber\\
&\approx(\bar{\lambda_{1}}-\bar{\lambda_{2}})+\frac{1}{3}(\bar{\lambda_{1}}-\bar{\lambda_{2}})^{3}\zeta^{2}\ \ (\zeta\ll1)\label{cal2}
\end{eqnarray}
We substitute the calculations of eqs. (\ref{cal1}) and (\ref{cal2}) 
into the first Bethe ansatz equation (\ref{eq:BAE11}) for the XXZ spin chain.
\begin{eqnarray}
2\tan^{-1}\Bigl(2\bar{\lambda_{1}}+\frac{1}{3}\bar{\lambda_{1}}^{3}\zeta^{2}\Bigr)=\frac{2\pi}{N}J_{1}+\frac{2}{N}\tan^{-1}\Bigl((\bar{\lambda_{1}}-\bar{\lambda_{2}})+\frac{1}{3}(\bar{\lambda_{1}}-\bar{\lambda_{2}})^{3}\zeta^{2}\Bigr).\nonumber\\
\label{BAE11corres}
\end{eqnarray}
It is clear that 
the Bethe ansatz equation (\ref{BAE11corres}) converges to that of the 
spin-1/2 XXX spin chain as we send parameter $\zeta$ to zero: 
$\zeta\rightarrow0$, if the reduced rapidities $\bar{\lambda_{1}}$ and $\bar{\lambda_{2}}$ do not diverge and remain finite. 
It has been shown rigorously in the two down-spin sector 
\cite{DG1} that all the solutions of the Bethe ansatz equations 
of the XXX spin chain are associated with the distinct sets of the 
Bethe quantum numbers $J_1^{\rm XXX}$ and $J_2^{\rm XXX}$, and also that  
the solutions are distinct. 
 Therefore, if $\zeta$ is small enough, then the reduced rapidities  
$\bar{\lambda}_1$ and $\bar{\lambda}_2$ are very close to 
the solution of the spin-1/2 XXX spin chain  whose 
Bethe quantum numbers $J_1^{\rm XXX}$ and $J_2^{\rm XXX}$ are 
given by $J_1$ and $J_2$, respectively, which are  
the Bethe quantum numbers of the XXZ solution, 
$\lambda_1$ and $\lambda_2$. 

Suppose that  there is a finite solution
$\tilde{\lambda}_1$ and $\tilde{\lambda}_2$  
of the XXX spin chain that has  the same set of the Bethe quantum 
numbers $J_1$ and $J_2$, i.e.,  $J_1^{\rm XXX}=J_1$ and $J_2^{\rm XXX}=J_2$. 
If the reduced rapidities $\bar{\lambda}_1$ and $\bar{\lambda}_2$ do not converge to  the XXX solution $\tilde{\lambda}_1$ and $\tilde{\lambda}_2$ in the limit of 
sending $\zeta$ to zero, then it is against the uniqueness of the 
XXX solutions with respect to the Bethe quantum numbers, 
which has been shown in Ref. \cite{DG1}. 

We therefore conclude that we have argued that for  any given finite solution of 
Bethe ansatz equations in the spin-1/2 XXX spin chain, there exists a 
solution of the Bethe ansatz equations in the spin-1/2 XXZ spin chain 
such that they have the same set of 
the Bethe quantum numbers in common and in the limit of sending 
$\zeta$ to zero the XXZ solution converges to the XXX solution.

\section{Conclusions} 

In this paper, by deriving all the Bethe quantum numbers for 
the complex solutions of the Bethe ansatz equations  in the two down-spin sector  
we have shown analytically the number of 
complex solutions in the massive regime of the spin-1/2 XXZ Heisenberg spin chain. We have derived the following results.  
(\si) We have shown the existence of extra two-string solutions, which are not 
described by the string hypothesis. 
We derived the criterion for extra two-string solutions to exist 
in terms of the anisotropy parameter $\Delta$ and the site number $N$. 
(\sii) We formulated  the criterion  for the collapse of $m$ two-string solutions occurs in terms of the anisotropy parameter $\Delta$ and the site number $N$, 
by assuming the monotonicity of the counting function. 
(\siii) In the stable regime we have shown rigorously that 
the counting function is monotone in $w$ for large $N$.   
In the unstable regime we confirmed the monotonicity numerically,   
and presented  analytical arguments for it. 
(\siv) 
We have shown that  string deviations are exponentially small 
as the site number $N$ increases for  two-strings with wide pairs 
or with narrow pairs in the stable regime 
for the two down-spin sector in the massive regime of  the spin 1/2 XXZ spin chain. 
(\sv) We presented  the numerical estimates of extra string solutions for 
$N=12$. We have illustrated the Bethe quantum numbers with the numerical solutions for $N=12$. 
(\svi) We have argued that for a finite-valued solution of the spin-1/2 XXX spin chain there exists  a solution of  the massive XXZ spin chain 
that has the same  Bethe quantum numbers in common.

\ack
We thank Chisa Hotta for useful comments. This work was partially supported 
JSPS KAKEN Grants No. JP18K03450.

\appendix
\section{Branches of logarithmic function}\label{sec:Branch}

\subsection{Symmetric branch with a cut on the negative $x$ axis}
We shall define  the arctangent function in terms of a logarithmic function defined for nonzero complex numbers $z=\alpha+i\beta$ where $\alpha$ and $\beta$ are real number. 
The logarithmic function $\log^{(s)} z$ is given by   
\begin{eqnarray}
\log^{(s)}(\alpha+i\beta) = i (\theta^{(s)}(\alpha+i\beta)+2\pi n)+\frac{1}{2}\log(\alpha^{2}+\beta^{2})\label{eq:A1}
\end{eqnarray}
where $n$ is an integer ($n\in \bf{Z}$) corresponding to the branch of the logarithmic function and we express the angle $\theta^{(s)}(z)$ as
\begin{eqnarray}
\theta^{(s)}(\alpha+i\beta)=\left\{ \begin{array}{ll}
\tan^{-1}(\beta/\alpha)+\pi H(-\alpha)\mbox{sgn}(\beta_{+}) & \mbox{for}\ \alpha\neq 0 \\
\mbox{sgn}(\beta_{+}) \pi/2 & \mbox{for}\ \alpha= 0 \\
\end{array} \right.\label{eq:A2}
\end{eqnarray}
Here we recall that we take the branch: $-\pi/2<\tan^{-1}x<\pi/2$ for ($x\in\bf{R}$). We denote by $\mbox{sgn}(x_{+})$ a sign function sgn(x) shifted by an infinitesimal positive number $0_{+}$: $\mbox{sgn}(x_{+})=\mbox{sgn}(x+0_{+})=1-2H(-x)$.

The function $\theta^{(s)}(\alpha+i\beta)$ defined by (\ref{eq:A2}) is continuous at $\alpha=0$ as a function of $\alpha$ when $\beta\neq 0$. If $\beta\neq 0$, the range is given by $-\pi<\theta(z)<\pi$. If $\beta=0$ and $\alpha>0$, $\theta(z)=0$. If $\beta=0$ and $\alpha<0$, $\theta(z)=\pi$.

\subsection{Positive branch with a cut on the positive $x$ axis}

The logarithmic function $\log^{(+)} z$ is given by   
\begin{eqnarray}
\log^{(+)}(\alpha+i\beta) = i (\theta^{(+)}(\alpha+i\beta)+2\pi n)+\frac{1}{2}\log(\alpha^{2}+\beta^{2})\label{eq:A3}
\end{eqnarray}
where $n$ is an integer ($n\in \bf{Z}$) corresponding to the branch of the logarithmic function and we express the angle $\theta^{(+)}(z)$ as
\begin{eqnarray}
\theta^{(+)}(\alpha+i\beta) = \left\{ \begin{array}{ll}
\tan^{-1}(\beta/\alpha)+\pi H(-\alpha) + 2 \pi H(\alpha_+) H(-\beta)  & \mbox{for}\ \alpha\neq 0 \\
\mbox{sgn}(\beta_{+}) \pi/2 + 2 \pi H(-\beta) & \mbox{for}\ \alpha= 0 \\
\end{array} \right. \nonumber \\
\label{eq:A4}
\end{eqnarray}

(The function $\theta^{(+)}(\alpha+i\beta)$ defined by (\ref{eq:A2}) is continuous at $\alpha=0$ as a function of $\alpha$ when $\beta\neq 0$. If $\beta\neq 0$, the range is given by $-\pi<\theta(z)<\pi$. If $\beta=0$ and $\alpha>0$, $\theta(z)=0$. If $\beta=0$ and $\alpha<0$, $\theta(z)=\pi$. )

\subsection{Arctangent function}

We define the arctangent function $\tan^{-1}(a+ib)$ for a nonzero complex number $a+ib$ where $a$ and $b$ are real by 
\begin{eqnarray}
\widehat{ \tan^{-1}}(a+ib)=\frac{1}{2i}(\log^{(+)}(1-b+ia)-\log^{(s)} (1+b-ia))\label{eq:A5}
\end{eqnarray}

Applying formulae (\ref{eq:A1})-(\ref{eq:A3}) We have for $b\neq\pm1$
\begin{eqnarray}
& & 2 \widehat{ \tan^{-1}}(a+ib)=\tan^{-1}\biggl(\frac{a}{1-b}\biggr)+\pi H(b-1)\mbox{sgn}(a_{+})\nonumber\\
&& \quad +2 \pi H(1-b) H(-a)  
+ \tan^{-1}\biggl(\frac{a}{1+b}\biggr) - \pi H(-b-1)\mbox{sgn}(a_{-})\nonumber\\
&& \quad +\frac{1}{2i}\log\biggl(\frac{a^{2}+(b-1)^{2}}{a^{2}+(b+1)^{2}}\biggr)\label{eq:A6}
\end{eqnarray}
Here we hajve made use of the fact that $b > 0$. 

In the branch: $-\pi<Im\log z\leq\pi$ we can show
\begin{eqnarray}
-\pi<&&Re (2\tan^{-1}(a+ib))<\pi\ \ (a\neq0),\nonumber\\
&&Re (2\tan^{-1}(a+ib))=\pm\pi\ \ (a=0)\label{eq:A5}
\end{eqnarray}

\section{Derivatives of $\widetilde{A}(w)$,  $\widehat{B}(w)$ and $\widehat{C}(w)$}
\label{sec:derivatives}

\subsection{Derivatives of $\widetilde{A}(w)$}
In the interval $0 < w <1/t$ for both narrow and wide pairs we have 
\begin{eqnarray} 
{\frac {d \widetilde{ A}} {d w}}(w) & = {\rm{sgn}}(1-w) \, 
2 t^2 \left( \frac {N_{\zeta}} N - 1  + w  \right) 
\left| \frac {1 + w t^2 } {1-w}  \right|^{1-{\frac 2 N}}  \nonumber \\ 
&  \quad + 2 t^2 \left( \frac {N_{\zeta}} N - 1  -  w  \right) \left( \frac {1 - w t^2 } {1+w}  \right)^{1-{\frac 2 N}}    
\end{eqnarray}
\begin{eqnarray}
\frac {d \widehat{ A}} {dw} & = \widetilde{ A} + w {\frac {d \widetilde{ A}} {d w}} 
\nonumber \\ 
& = - 2 t^2 \frac {N_{\zeta}} N w \left( 
 {\rm{sgn}}(w-1) \, 
 \left| \frac {1 + w t^2 } {w-1}  \right|^{1-{\frac 2 N}} 
-  \left( \frac {1 - w t^2 } {1+w}  \right)^{1-{\frac 2 N}}    
\right) \nonumber \\ 
& + (1-3 w t^2) (1-wt^2)^{1-2/N} (w+1)^{2/N} \nonumber \\ 
& - (1 + 3 w t^2) (1 + wt^2)^{1-2/N} |w-1|^{2/N} \, . 
\end{eqnarray}

\subsection{Derivative of $\widehat{B}(w)$}
In the interval $0 < w <1/t$ for both narrow and wide pairs we have 
\begin{eqnarray} 
\frac {d \widehat{B}} {d w} & = 
\frac   {(1 -w^2 t^2) (1 + 3 w^2 t^2)} {t^2 w^2} 
\Bigg\{ \left( \frac {1+w} {1-wt^2} \right)^{2/N} 
- \left( \frac {1-w} {1+wt^2} \right)^{2/N} \Bigg\} \nonumber \\ 
& + 
\left| \frac {1-w} {1+wt^2} \right|^{2/N}   
\, 
\Big\{ - {\frac {2 N_{\zeta}} N}  \frac  {(1- w^2 t^2)^2} {w(1-w)(1+wt^2)} 
- {\frac {4 (1+t^2)} N}  {\frac {1+w} {1-w} } \nonumber \\ 
& \qquad + 2 (1+t^2 +2 t^2 w) \Big\} 
\nonumber \\
& +  \left( \frac {1+w} {1-wt^2} \right)^{2/N} 
\Big\{- {\frac {2 N_{\zeta}} N}  \frac {(1- w^2 t^2)^2} {w (1+w)(1-wt^2)}  
+ {\frac {4 (1+t^2)} N}  {\frac {1-w} {1+w} } 
\nonumber \\ 
& \qquad - 2 (1+t^2 - 2t^2 w ) \Big\} \, . \label{eq:dBhat} 
\end{eqnarray} 

\subsection{Derivative of $\widehat{C}$}
In the interval $0 < w <1/t$ for both narrow and wide pairs we have 
\begin{eqnarray} 
\frac {d \widehat{C}} {d w} & = 
 \left| \frac {1-w} {1+wt^2} \right|^{2/N} 
\left\{ - \frac {1 -w^2 } {w^2} - {\frac {2 N_{\zeta}} N}  {\frac {t^2} w} 
\frac {(1+w)^2} {(1-w)(1+wt^2)} \right\}
\nonumber \\  
& + \left( \frac {1+w} {1-wt^2} \right)^{2/N} 
\left\{ \frac {1 -w^2 } {w^2} 
- {\frac {2 N_{\zeta}} N}  {\frac {t^2} w}  
\frac {(1-w)^2} {(1+w)(1-wt^2)} 
\right\} .  \label{eq:dChat} 
\end{eqnarray}

\section{Useful expansion}\label{sec:expansion} 

For $0 < w <1$ we have 
\begin{eqnarray}
&\left( \frac {1-w} {1+w t^2}  \right)^{2/N} = 
1 - {\frac {2(1+t^2)} N} w + \left( - {\frac {1-t^4} N} + \frac {2(1+t^2)^2} {N^2}  \right) w^2 
\nonumber \\ 
& \quad + \left( - \frac {2 (1+t^6)} {3N} +  \frac {2(1+t^2)(1-t^4)} {N^2} - 
\frac {4 (1+t^2)^3} {3 N^3}  \right) w^3 + \cdots . \label{eq:exp}
\end{eqnarray}

\section{Takahashi quantum numbers }\label{sec:expansion}

In the string hypothesis two-strings are associated with 
the Takahashi quantum number $I$. For two-string solutions 
of the spin-1/2 XXX spin chain with $N$ sites in the sector of two down-spins 
the Takahashi quantum number $I$ is given by 
\begin{eqnarray}
I = -N/2 +2, -N/2 + 3 , \cdots, N/2 -2. 
\end{eqnarray}   
It has been shown \cite{DG1} that the 
Takahashi 
quantum number $I$ is 
related to the Bethe quantum numbers $J_1$ and $J_2$ 
for the two-string solutions of the spin-1/2 XXX spin chain 
in the sector of two down-spins as follows. 
\begin{equation}
I = J_1 + J_2 - \frac N 2 {\rm sgn}(x) \, . 
\end{equation}
Here, $x$ denotes the center of two strings. 
in the XXX chain two string solutions are given by 
$\lambda_1 = x + i/2 + i \delta$ and  $\lambda_2 = x - i/2 - i \delta$.  
 
For wide pairs, since $J_2=J_1+1$, 
the largest Bethe quantum number $J_1^{\rm max}$ satisfies 
\begin{equation}
2 J_1^{\rm max} +1 -N/2 = N/2 - 2 \, . 
\end{equation}
It follows that we have $J_1^{max}= (N-3)/2$. 
For narrow pairs, since $J_2=J_1$ the largest Bethe quantum number $J_1^{\rm max}$ satisfies 
\begin{equation}
2 J_1^{\rm max}  -N/2  \le  N/2 - 2 \, . 
\end{equation}
We have 
\begin{equation}
J_1^{\rm max} \le (N-2)/2 \, . 
\end{equation}
From the parity condition 
we have $J_1^{\rm max}= (N-3)/2$.

\end{document}